\documentclass[apj]{emulateapj}  

\shorttitle{The most ancient spiral galaxy} 
\shortauthors{Yuan et al.}
\journalinfo{ApJ accepted 2017 October 18}
\submitted{ApJ accepted 2017 October 18}

\usepackage{color}
\usepackage[colorlinks=true,
            linkcolor=black,
            urlcolor=black,
            citecolor=blue, breaklinks=true]{hyperref}
\usepackage{url}
\usepackage{breakurl}

\usepackage{amsmath}
\usepackage{amssymb}
\usepackage{gensymb}

\newcommand{\ergscmarc}{{erg\,s$^{-1}$\,cm$^{-2}$\,arcsec$^{-2}$}}

\newcommand{\kms}{{km\,s$^{-1}$}}

\newcommand{\ms}{$M_{\rm star}$}

\newcommand{\ha}{{H$\alpha$}}
\newcommand{\nii}{{[N\,{\sc ii}]}}

\newcommand{\lam}{$\,\lambda$}

\begin{document}


\title{The most ancient spiral galaxy:  a 2.6-Gyr-old disk with a tranquil velocity field}


\author{Tiantian Yuan\altaffilmark{1,2,4}\altaffilmark{*}, Johan Richard\altaffilmark{3}, Anshu Gupta\altaffilmark{4}, Christoph Federrath\altaffilmark{2,4}, Soniya Sharma\altaffilmark{4},  Brent A. Groves\altaffilmark{2,4}, Lisa J. Kewley\altaffilmark{2,4}, Renyue Cen\altaffilmark{5}, Yuval Birnboim\altaffilmark{4,6}, David B. Fisher\altaffilmark{1}}
\affil{\scriptsize 
$^1$Centre for Astrophysics and Supercomputing, Swinburne University of Technology, Hawthorn, Victoria 3122, Australia; \\
$^2$ARC Centre of Excellence for All Sky Astrophysics in 3 Dimensions (ASTRO 3D), Australia; \\
$^3$Univ Lyon, Univ Lyon1, Ens de Lyon, CNRS, Centre de Recherche Astrophysique de Lyon UMR5574, F-69230, Saint-Genis-Laval, France;\\
$^4$Research School of Astronomy and Astrophysics, The Australian National University, Cotter Road, ACT 2611, Australia;\\
$^5$Princeton University Observatory, Princeton, NJ 08544, USA;\\
$^6$Racah Institute of Physics, The Hebrew University, Jerusalem Israel
}
\altaffiltext{*}{ASTRO 3D Fellow}
\email{Email: tiantianyuan@swin.edu.au}

\begin{abstract}
We report an integral-field spectroscopic (IFS) observation of a gravitationally lensed spiral  galaxy {\it A1689B11} at  redshift $z=2.54$. It is the most ancient spiral galaxy discovered to date and the second kinematically confirmed spiral  at $z\gtrsim2$.  Thanks to gravitational lensing, this is also by far the deepest IFS observation
   with the highest spatial resolution ($\sim$ 400 pc) on a spiral galaxy at a cosmic time when the Hubble sequence is about to emerge.    After correcting for a lensing magnification of 7.2 $\pm$ 0.8,  this  primitive spiral disk has an intrinsic star formation rate of 22 $\pm$ 2 $M_{\odot}$ yr$^{-1}$,  a stellar mass of 10$^{9.8 \pm 0.3}$$M_{\odot}$ and a half-light radius of $r_{1/2}=2.6 \pm 0.7$ kpc,  typical of a main-sequence  star-forming (SF) galaxy at $z\sim2$.  However, the \ha\ kinematics show a surprisingly tranquil velocity field with an ordered rotation  ($V_{\rm c}$ {=} 200 $\pm$ 12 \kms) and  uniformly small velocity dispersions ($V_{\rm \sigma, mean}$ {=} 23 $\pm$ 4 \kms and $V_{\rm \sigma, outer-disk}$ {=} 15 $\pm$ 2 \kms).     The low gas velocity dispersion is similar to local spiral galaxies and is consistent with  the classic density wave theory where  spiral arms form in dynamically cold and thin disks.      We speculate that {\it A1689B11} belongs to a population of rare spiral galaxies at $z\gtrsim2$ that mark the formation epoch of thin disks. Future observations with JWST 
   will greatly increase the sample of these rare galaxies and unveil  the earliest onset of spiral arms.
 \end{abstract}


\section{Introduction}
\label{sec:intro}
One of the most common features of disk galaxies in the local universe is the presence of prominent spiral arms.  Among millions of galaxies charted
 in the local universe, $\sim$70\% exhibit spiral arms \citep[e.g.,][]{Nair10,Willett13}.  However, the number density  of spiral galaxies decreases dramatically at high redshift \citep{Conselice14,ElmegreenB06}.  For example, only one spiral galaxy has been spectroscopically confirmed at $z\gtrsim2$ \citep{Law12n}.  

Spiral arms serve important purposes in galaxy formation and evolution: they are sites of star formation and are intimately associated with the formation of the thin and thick disk \citep{ElmegreenB11,Conselice14,MartinezMedina15}.  Spiral arms  play an active role in driving the radial and azimuthal mixing of the metals, redistributing angular momentum,  and smoothing out small-scale mass distributions \citep[e.g.,][]{Sellwood02,Sellwood14,Grand15,Grand16}.  The number and pitch angle of spiral arms are 
strongly correlated with the  mass distribution of the disk and can be a powerful tool to constrain the bulge and black hole masses
 \citep{Athanassoula87,Kennicutt81,ElmegreenD90,Berrier13,Dobbs14r,Seigar14,Davis15,Davis17}. The onset of spiral structures offers crucial insights into the origin of the Hubble sequence \citep{Driver98, Cen14a,Genel15}.

The necessary and sufficient conditions for spiral arm formation remain inconclusive, despite  major developments in the 1960s  and decades of studies \citep[e.g.,][]{Toomre77,Athanassoula84,Sellwood11,Dobbs14r}.  Popular mechanisms for spiral arm formation are largely based on early analytical works: e.g., the density wave theory \citep{Lindblad60,Lin64,Kalnajs71}, swing amplifications \citep{Goldreich65,Julian66}, and bars and tidal interactions \citep{Kormendy79,Salo93}.   These three mechanisms are not 
mutually exclusive and have mixed observational  successes \citep[e.g.,][]{Sellwood11,DOnghia13,Shu16,PourImani16}.  Whether spiral arms are long-lived patterns or transient features is still hotly debated in theory and poorly constrained in observations  \citep{Sellwood02,Sellwood11}. The progress in the theory of spiral arm formation is slow and current efforts primarily focus on nearby galaxies \citep{Dobbs14r}.

Breakthroughs can come from observations of high-redshift galaxies when spiral arms are in the early stages of formation. 
All classic  analytical models of spiral arm formation assume an infinitesimally thin and cold disk in a stable rotation \citep[e.g.,][]{Toomre77,Bertin96,Rafikov01,Sellwood14}.   
It is unclear if this assumption holds at high redshift. Compared to local galaxies, high-redshift disks are  gas-rich, globally unstable, and tend to have larger velocity dispersions  and thicker disks \citep[e.g.,][]{Freeman02,ElmegreenB06, Law07,Tacconi13,Glazebrook13r,Wisnioski15,Johnson17,Zhou17}.   
The role of the interstellar medium (ISM) and gas feedback in shaping spiral arms  is usually  over-simplified  in local galaxies, but could become increasingly important and complicated at high redshift \citep[e.g.,][]{Bertin88,ElmegreenB93,Wada11,Ghosh15}. 
Spiral arms are also sensitive to  external   processes such as bars, galaxy mergers,  and gas accretion, all of which are  different at high redshift   \citep{Sellwood04,Wada11,Martig12,Dobbs14r}.     Studying spiral galaxies at a dynamically hostile cosmic time 
 has the unique advantage of probing the most sensitive factor(s) responsible for  spiral arm formation.

Spiral galaxies are rarely observed at $z\gtrsim2$  \citep{ElmegreenB06,Law12n,Conselice14}. A minor-merger triggered face-on spiral  at $z=2.18$ remains the only  thoroughly studied case in the literature  \citep{Law12n}.  Previous  data show that  spiral arms are less well-developed and more chaotic beyond $z=0.5$ \citep{Abraham01}. The onset of spiral structures in galaxies is proposed to occur at $z\sim1.8$, when disks have  developed a cool stellar component in a rotation-dominated disk \citep{ElmegreenD14}.  However, indirect observational evidence shows that the Hubble sequence may already be in place at $z\sim2.5$ \citep{Wuyts_S11}, implying an even earlier onset  of spiral arms. 

 Observational difficulties in identifying spiral arms at high redshift might have contributed to the rarity of spiral galaxies at $z\gtrsim2$.
 In the local universe,  spiral arms are visually classified through qualitative inspection of morphologies \citep[e.g.,][]{
Hubble26,Reynolds27,Sandage05}.  This visual classification scheme suffers strongly  from observational biases caused by degraded resolution, cosmological surface brightness dimming, band shifting and imaging depth at high redshift \citep{Abraham96,Giavalisco96,Hibbard97,Conselice00,Conselice11}.  For example, spiral features can only be reliably identified for $\sim270$ galaxies in the Hubble Ultra Deep Field (UDF)  that have major axes larger than 10 pixels  \citep{ElmegreenD05}. Most galaxies at $z\gtrsim2$ have half-light radii of $\le 0.\!\!^{\prime\prime}3$ \citep{Allen17},  making the identification and quantification of spiral features such as   the number of arms and pitch angles challenging \citep{Shields15}.
More studies like \citet{van-den-Bergh02} focusing on testing the visibility of spiral arms  at high redshift are required before concluding  the  actual number density of spiral galaxies at   $z > 1$.

In addition to observational biases,  cosmological simulations suggest a few physical processes at $z\gtrsim2$ that hinder the formation of spiral arms.   
Earlier cosmological simulations suggest that the paucity of spiral galaxies at $z\gtrsim2$ can be ascribed to
  high merger rates  \citep{Hammer09},  high gas accretion rates and multiple low angular momentum inflow cold streams \citep{Cen14a}. 
Some cosmological simulations report that grand-design spirals are in place by $z\sim$ 3 \citep{Fiacconi15} and the high-redshift spirals 
most likely originated from swing amplifications  triggered by satellites.  Detailed observations of spiral galaxies at high redshift will enable meaningful investigations into these physical processes that remain elusive in simulations.

Gravitationally lensed galaxies with adaptive-optics (AO) aided observations have pioneered the measurement of physical properties of high-redshift galaxies \citep[e.g.,][]{Swinbank07b,Stark08,Jones10b,Yuan11,Yuan12,Yuan15,Swinbank15}. The lensing magnification  allows selection of less massive systems
and measurements on smaller physical scales  than  magnitude-limited studies.  The spatial resolution in gravitational lensing observations
can reach a few times 10-100 pc, important for minimizing beam-smearing effect and resolving star clusters \citep[e.g.,][]{Jones10a, Yuan13b, Livermore15,Vanzella17,Rigby17}.   In this paper, we report  the integral-field spectroscopic (IFS)  observation of a gravitationally lensed spiral galaxy {\it A1689B11} at $z=2.54$.  It is the second spectroscopically  confirmed spiral galaxy at $z\gtrsim2$ and is  $\sim$10 times less massive than the spiral galaxy of \citet{Law12n}.

 This paper is organized as follows.
 Section~\ref{sec:data} describes our IFS observation,  data analysis and lens models. 
 Section~\ref{sec:results} describes our results from the IFS data and derived physical properties of  {\it A1689B11}.
 In Section~\ref{sec:disc} we discuss the nature of {\it A1689B11} and compare it to other galaxy samples. 
We summarize and conclude in Section~\ref{sec:future}.     Throughout this paper, we adopt a standard $\Lambda$CDM cosmology with $\Omega_{M}{=}0.3$, $\Omega_{\Lambda}{=}0.7$ and H$_{0}{=}70$~km~s$^{-1}$~Mpc$^{-1}$.  At the redshift of  $z=2.54$,
1$\arcsec$ corresponds to a physical scale of 8.2 kpc.

\section{Observations and Data Reduction}\label{sec:data}
\subsection{Spiral Galaxy Candiate {\it A1689B11}}
\begin{figure*}[!ht]
\begin{center}
\includegraphics[trim = 0mm -10mm 5mm 0mm, clip, width=0.95\textwidth, angle=0]{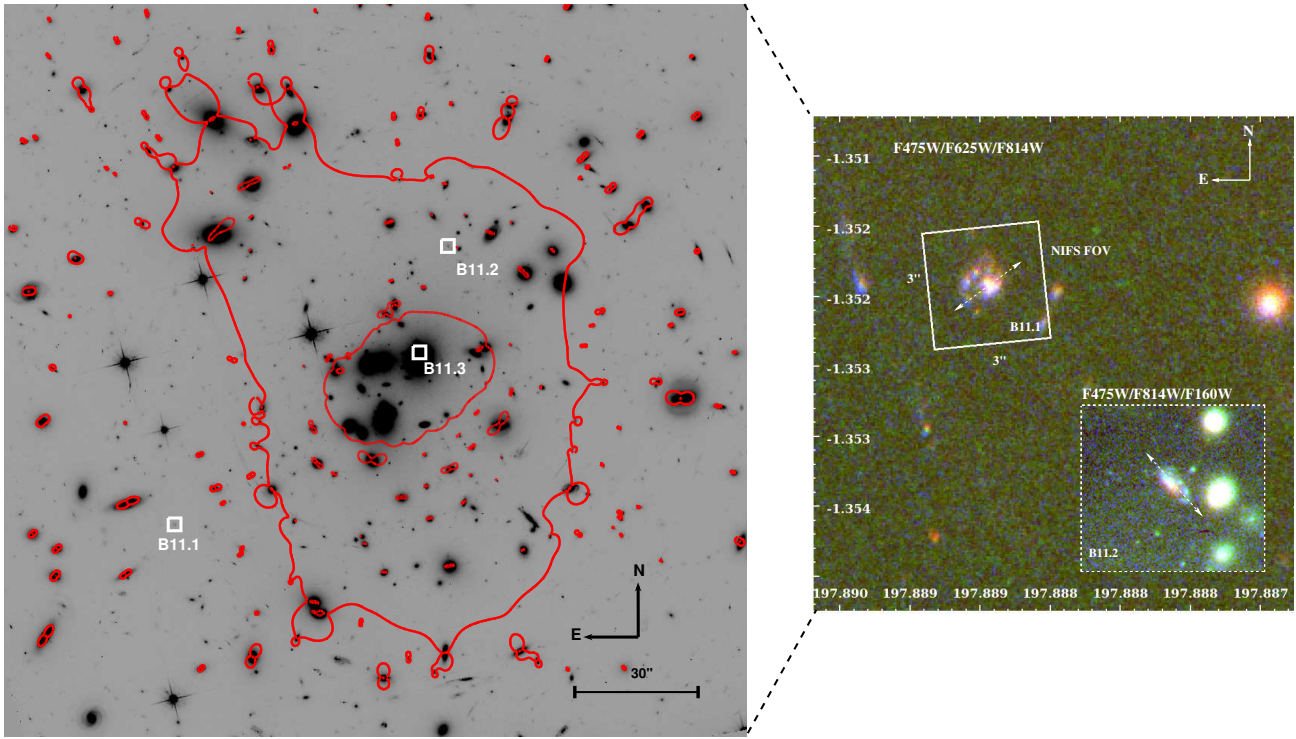}
\caption{Left:  The HST F814W image of the  strong lensing cluster Abell 1689. 
Red lines show the critical lines at  {\it A1689B11}'s  redshift of $z=2.54$ based on the lens model of \citet{Limousin07}.  The outer red line is the tangential critical line; the inner red line is the radial critical line.
White boxes show the positions of  three multiple lensed images of  {\it A1689B11}.   B11.1 and B11.2 are well-resolved magnified images; B11.3 is the demagnified central image overlapping with  one of the brightest cluster members.   Right: The zoom-in HST 3-color (F475W/F625W/F814W) image (B11.1) of the lensed spiral galaxy  {\it A1689B11}.  
The embedded 3-color (F475W/F814W/F160W) image (B11.2) shows extra color information from the HST WFC3 IR data. 
The dashed arrow denotes the major axis of the galaxy.  B11.1 is  stretched more in the minor-axis  direction whereas B11.2 is stretched roughly equally in all directions by lensing.   B11.1 and B11.2 show identical spiral morphology on both the image plane and the source plane in all available HST broad-band images (Appendix).   Our NIFS/Gemini observation (solid white box) is centered on image B11.1 because of the laser guide star requirements.   
\label{fig:fig1}}
\end{center}
\end{figure*}

The spiral galaxy candidate (hereafter {\it A1689B11}) was first recorded as ``source 11" with a 
photometric redshift of $z=2.9 \pm 0.2$  
in the strong-lensing analysis of the galaxy cluster Abell 1689 \citep{Broadhurst05}.
Source 11 is gravitationally lensed into
 two highly magnified images B11.1 and B11.2, and a central image B11.3 (Figure~\ref{fig:fig1}).   \citet{Broadhurst05}
 was the first to point out the spiral feature of {\it A1689B11} and speculated it would be the highest redshift spiral galaxy
 if confirmed.   The first spectroscopic redshift of  $z=2.5$ was provided in \citet{Limousin07} based on rest-frame UV spectroscopy with Keck/LRIS. 
The detection of   \ha\ lines  at $z=2.54$ was reported in the near-infrared (NIR) multi-slit 
survey of  \citet{Yuan13a}.   This work presents the NIR IFS observation of image B11.1 (Figure~\ref{fig:fig1}).
The  imaging and photometric data used in this work are  obtained from the Hubble Space Telescope (HST; proposal IDs: 11802, 9289, 11710)  and 
the Spitzer Space Telescope archives  (program ID: 20439).

 \subsection{NIFS/Gemini Observations and Data Reduction}\label{subsec:nifs}
AO aided NIR integral field spectroscopic observations with NIFS \citep[Near-infrared Integral-Field Spectrograph;][]{McGregor03} 
 were conducted between March 2013 and March 2014  on the Gemini North telescope  under excellent weather conditions (average seeing $\sim$  0\farcs5, airmass 1-1.6).    Our total allocated observational time was  13.5hrs in band 1 (program ID: GN-2013A-Q-23-64). 
  The observation implemented a dithering pattern of ``ABAABA" or  ``AB", i.e., $\sim$ 33-50\% of the target exposure time was spent on sampling sky frames in order to facilitate a good sky background subtraction.    The observation centered on image B11.1 instead of B11.2 because of the laser-guide star requirement of NIFS.    The coordinates of the pointing center are given in Table~\ref{tab:tab1}. The field of view of NIFS is 3\farcs0 $\times$  3\farcs0, with 29 slitlets, each  0\farcs1 wide.   NIFS  delivers a spectral resolving power of R $\sim$ 5300 in the K band, corresponding to a  rest-frame Gaussian velocity resolution of   $\sigma$  $\sim$ 24 \kms.

The data were reduced using a varied  Gemini IRAF package following standard procedures \citep[e.g.,][]{StorchiBergmann09}.  
This IRAF package produces sky-subtracted, telluric-corrected and flux-calibrated datacubes.  
Each individual 900s exposure  was spatially aligned based on the brightest \ha\ spaxels and  co-added based on a mean sigma-clipping 
procedure.  The 1$\sigma$ error datacubes were generated during the sigma-clipping process.   We used the  star HIP67004  of  spectral type A0V for telluric correction and flux calibration.   The systematic uncertainties in the flux calibration is estimated to be within 20\%.  A total of  $27\,\mathrm{ks}$ on-source  exposures were obtained in the K-band with an angular resolution  of $\sim$ 0.1\arcsec, corresponding to a median physical scale of $\sim$ 400 pc on the source plane of {\it A1689B11}.  Our observation reached  a 3$\sigma$ \ha\ emission line surface brightness depth of 3$\times$10$^{-17}$ \ergscmarc.

The astrometry of the datacube is calibrated  by  assigning  the HST coordinate of the galaxy center to the brightest spaxel in the wavelength-collapsed  2-dimensional (2D) NIFS image and rotating with the positional angle from the observation. The astrometric uncertainty  of our IFS datacube is about one spaxel, i.e., $\sim$ 0.$\!\!^{\prime\prime}$1.

\begin{figure*}[!ht]
\centerline{
\includegraphics[trim = 5mm 15mm 135mm 8mm, clip, width=0.9\textwidth,angle=0]{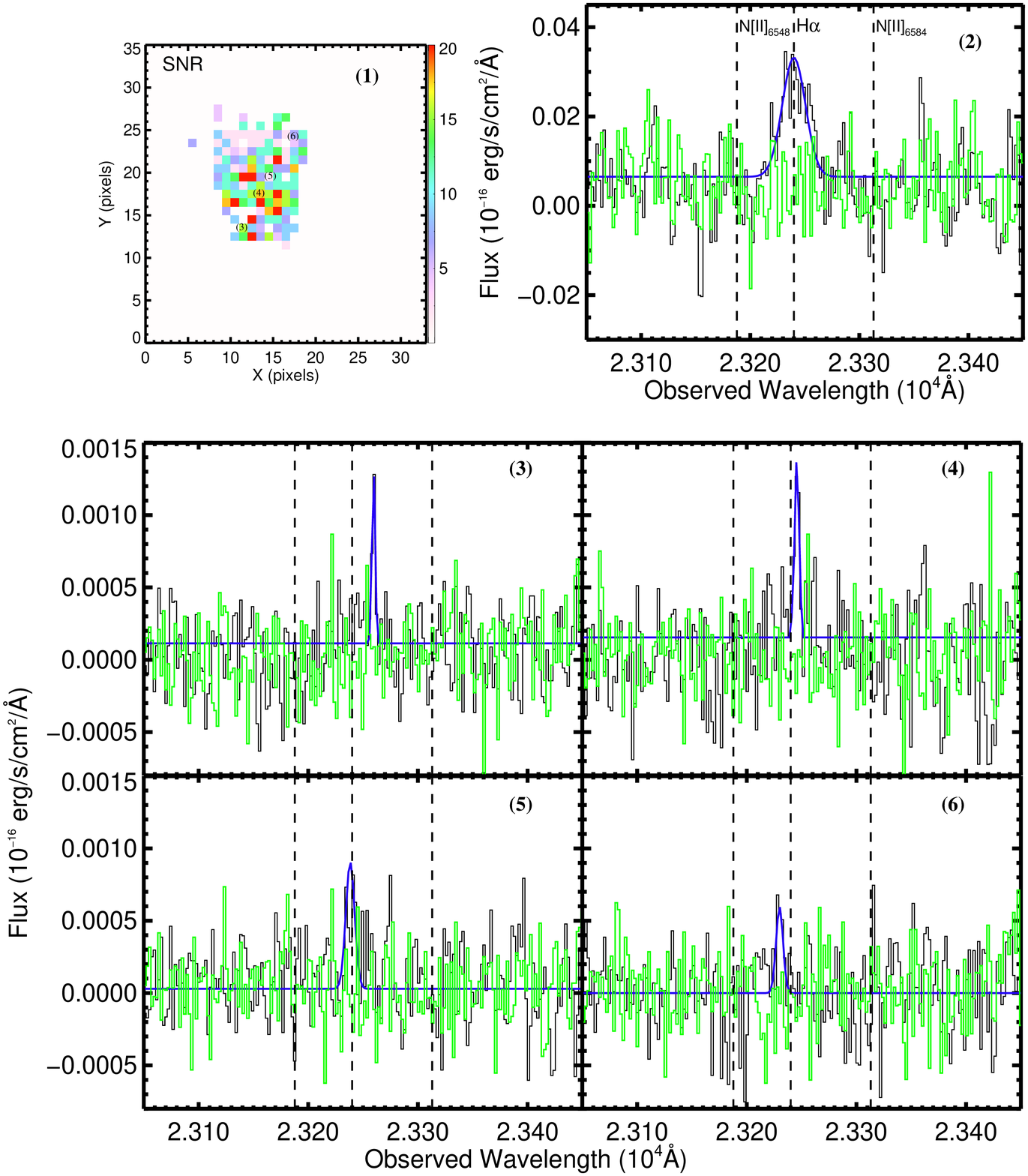}
}
\caption{Examples of NIFS 1D spectra and 2D SNR map in the observed frame.  Panel (1): 2D SNR map from \ha\ emission line fitting (1 pixel corresponds to 0.$\!\!^{\prime\prime}$1).  Panel (2):  Integrated 1D spectrum from coadding all  NIFS spaxels, weighted by SNR.  Panel (3)-(6):  1D spectrum from individual spaxels, chosen along the major axis of the galaxy to represent a range of SNR and velocity centroids. The locations of the corresponding spaxels are mark on panel (1). 
For panels (2)-(6), raw spectra are shown as  black lines; the RMS of the sky residuals are shown in green;  the best-fit Gaussian profiles for \ha\ lines are shown in blue;
vertical dashes lines indicate the expected positions of the \nii\lam6548, \ha\lam6583, and \nii\lam6583 emission lines at the kinematic center of the galaxy.  \nii\ lines are neither detected in any spaxel nor the integrated 1D spectrum. 
This figure highlights that  \ha\ emission lines from individual spaxels are genuinely  narrow,  even without  beam-smearing corrections.  The widths of the emission lines are  approaching the instrumental
resolution of NIFS ($\sim$2.5 wavelength channels here). 
\label{fig:fig2}}
\end{figure*}

\subsection{Emission Line Fitting}\label{subsec:linefit}
Our K-band observation was originally planned to detect \ha\ and \nii\ lines.   To analyze the IFS spectra, we first collapse the datacube in the wavelength dimension in the vicinity of the \ha\ line to create an \ha\ 2D map. We use the raw \ha\ 2D map to generate an initial 2D mask that flags spaxels with  no obvious  \ha\ line detections.  We then manually inspect the  spectrum of each  individual spaxel and refine the mask with three types of visual flagging: (1) significant  emission lines, (2) possible emission lines, and (3) no obvious emission lines.   The visual inspection of the datacube is  necessary in order to reject spaxels that are spurious for low signal-to-noise ratio (SNR) data.  We use this visual mask to supervise our subsequent automatic emission line fitting procedure, i.e., we  demand the fitting result to be consistent with our visual mask.   For example, an emission line fitting result of SNR $\ge$ 5 should have a visual flag of (1), whereas a line detection of SNR $< 3$ should be consistent with a flag type of (3).  Those with $3\le$ SNR $< 5$ should match the flag type of (2). 

Our automatic emission line fitting procedure involves fitting Gaussian profiles simultaneously to  three emission lines in each spaxel:  \nii\lam6548, 6583 and \ha\ (Figure~\ref{fig:fig2}).    The line profile fitting is conducted using a $\chi^2$ minimization procedure weighted by the inverse of the variance spectrum.
The fitting result is very sensitive to the weighting used.  The 1$\sigma$ error datacube generated from the mean sigma-clipping procedure described in Section 2.2 does not provide  good weighting to the emission line fitting, as it results in the SNR of the emission line being over-estimated and inconsistent with our visual mask flags.  
Instead we use the variance over a sky region devoid of emission lines  as the weighting.  
We  select a range of sky spaxels to generate a variance spectrum  (green lines in Figure~\ref{fig:fig2}) for each spaxel using a bootstrap procedure.
The SNR of the emission line fitting result is consistent with the visual mask.
  This weighting method is similar to that used in \citet{Leethochawalit16} and is reasonable because the variance of the NIR data is usually dominated by sky residuals.

\ha\ lines are detected in $\sim$ 130 individual spaxels   at  $\ge$ 5$\sigma$ level (Figure~\ref{fig:fig2}, (1)).   \nii\ lines are not  detected above 3$\sigma$ in either  individual spaxels or the integrated spectrum (Figure~\ref{fig:fig2}, (2)). \citet{Yuan13a} report an \nii\  line detection on one of the four slits configured with different positional angles on B11.1 and B11.2, implying a spatial variation of metallicities.  We extract a mock slit spectrum from our NIFS data based on the slit setup of \citet{Yuan13a},
 but no \nii\ line is detected above 3$\sigma$.  Deeper IFS data from our ongoing OSIRIS/Keck observation on image B11.2 will help to discern \nii\ lines and  the spatial metallicity distribution of  {\it A1689B11}.   Figure~\ref{fig:fig2} shows the \ha\ 2D  SNR map and examples of  single Gaussian fits to  \ha\ emission lines of  individual spaxels.  
An integrated spectrum from coadded spaxels is also presented in Figure~\ref{fig:fig2} for comparison.
 Figure~\ref{fig:fig2}  highlights the necessity of high-spatial resolution in order to distinguish beam-smearing from intrinsic line width. It is also apparent from
  Figure~\ref{fig:fig2}  that the \ha\ lines detected on individual spaxels have  very narrow widths (approaching the NIFS instrumental line width)   and show systematic 
offsets in the centroids.  The focus of this paper is to report the kinematics of \ha\ lines (see Section~\ref{sec:results}).

\subsection{Gravitational Lens Modeling}\label{subsec:lensmodel}
Abell 1689 is one of the most studied lensing clusters with well-constrained mass models.  We use the best-fit  model from  \citet{Limousin07} and 
the   software {\tt Lenstool}\footnote{\tt https://projets.lam.fr/projects/lenstool/wiki} \citep{Kneib93, Jullo07} to reconstruct the source-plane properties of {\it A1689B11}.  

We show the lensing configuration and critical lines of {\it A1689B11}  in Figure~\ref{fig:fig1} (left).  
{\it A1689B11}  is lensed into a three-image system: B11.1, B11.2 and B11.3.  Image B11.2 lies in between the tangential
and the radial critical line, resulting in a large flux magnification factor ($\mu= 12 \pm 2$) and a relatively  undistorted image (i.e., equally magnified in  the major and  the minor axis).  Our NIFS observation is centered on the  less magnified ($\mu= 7.2 \pm 0.8$) image B11.1.  B11.1 is magnified more in the minor axis direction, making it appear more face-on than the intrinsic image.    The geometric magnification for image B11.1 is  $\sim$  4 (along the minor axis) $\times$ 1.8 (along the major axis).   
A demagnified central image B11.3 is predicted by the lens model and also observed  to be overlapping with one of the brightest cluster members of Abell 1689 \citep{Broadhurst05,Limousin07}.

 The statistical error of the lens modeling  is estimated from a set of 
Markov Chain Monte Carlo (MCMC) realizations implemented in {\tt Lenstool} and is typically $\sim$ 10\% for Abell 1689. Both multiple 
 images (B11.1 and B11.2) yield consistent  source-plane morphologies and positions within 0.2\arcsec.  Because B11.1 and B11.2 are not close to the critical lines,  
the systematic errors in the source-plane morphologies are small compared to the case of giant arcs. 
Our source-plane reconstructed morphologies are robust within the model statistical  uncertainties.    The NIFS datacube is reprojected to the source plane after being remapped  to the  HST  coordinates using the calibrated astrometry.  Emission line fitting analyses have been carried out on both the image  and source plane and the results are self-consistent.   We present both the image-plane and source-plane properties in the following sections.

\begin{figure*}[!ht]
\centerline{
\includegraphics[trim = 30mm 15mm 10mm 20mm, clip, width=0.96\textwidth,angle=0]{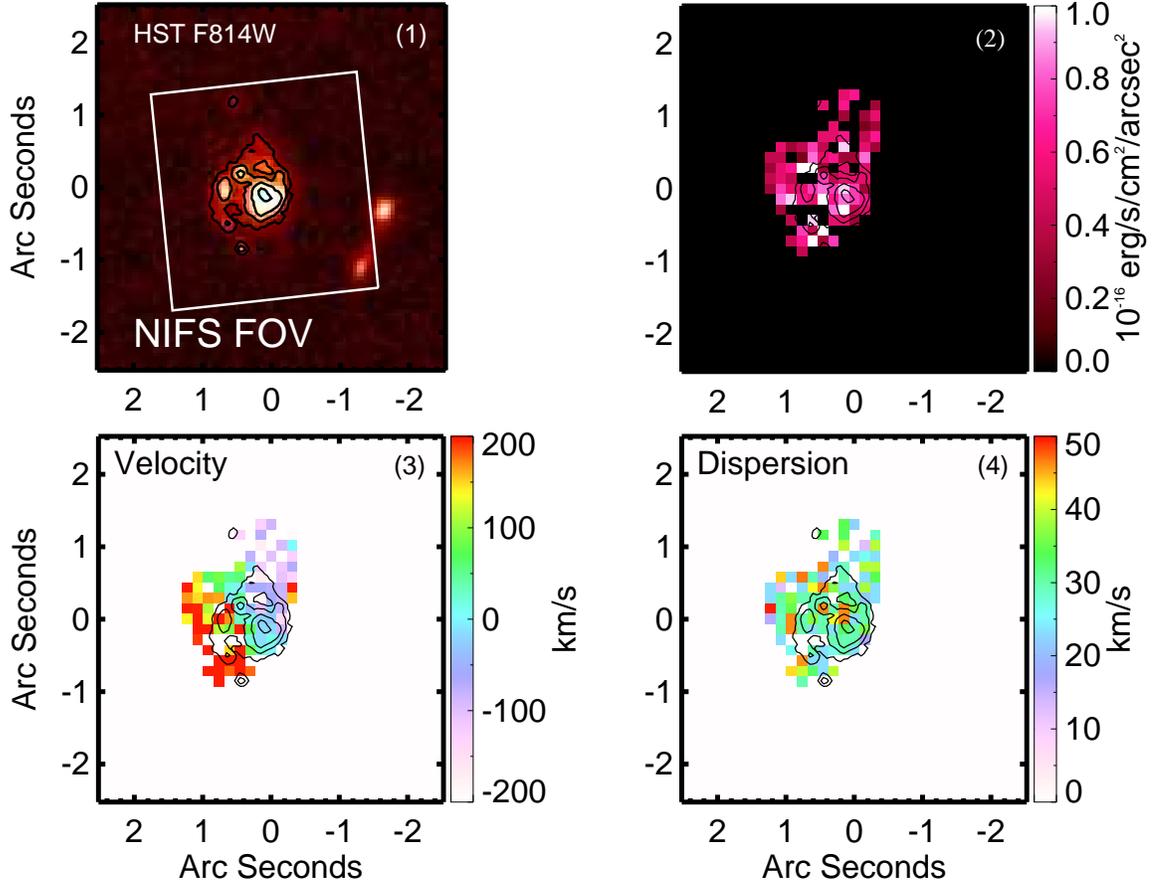}
}
\caption{HST morphology and NIFS 2D maps on the image-plane.
Panel (1): The HST F814W morphology of observed image B11.1. Black contours outline different surface brightness levels of the F814W image, including  star forming
clumps in the core and on the spiral arms.  The white box shows the NIFS FOV (3 $\!\!^{\prime\prime}$ by 3 $\!\!^{\prime\prime}$).   
(2): NIFS \ha\ intensity 2D map. 
(3): NIFS \ha\ velocity 2D map.
(4): NIFS \ha\ velocity dispersion 2D  map.
Black contours in (2)-(4) are the same as panel (1) and have been astrometrically aligned with the NIFS observation. 
All NIFS 2D maps are presented with the observed spaxel scale without smoothing/binning.  Only data with SNR $\ge$ 5 are included.
\label{fig:fig3}}
\end{figure*}

\begin{figure*}[!ht]
\centerline{
\includegraphics[trim = 10mm 10mm 15mm 0mm, clip, width=0.68\textwidth,angle=90]{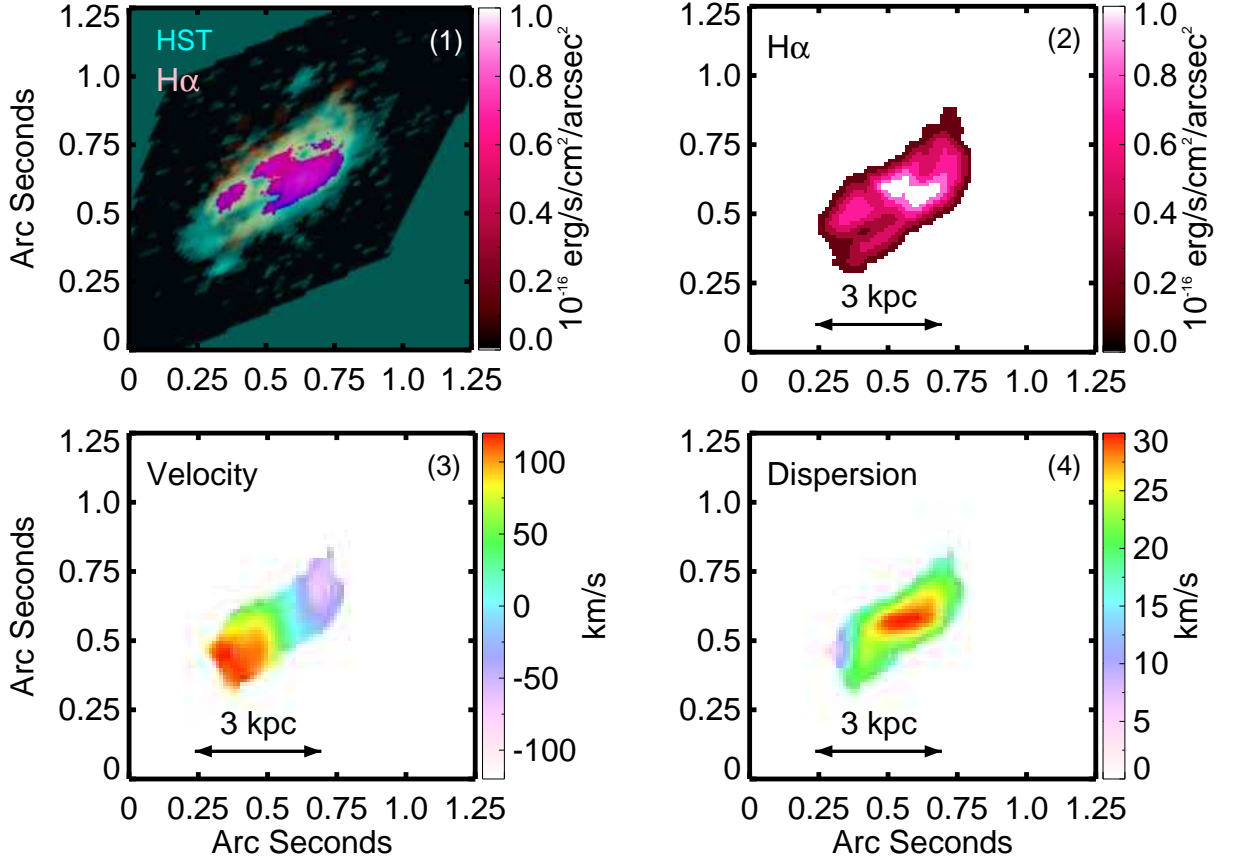}
}
\caption{The HST morphology and NIFS 2D maps on the source-plane.
Panel (1):   The NIFS \ha\ image (pink) on top of the HST F814W source-plane image (cyan). 
For the HST image, we used a subsampling of 20 on the image-plane and 10 on the source-plane to optimize the spatial resolution of the source-plane reconstruction.
The pixel scale of the source-plane HST image is therefore  0.$\!\!^{\prime\prime}$005. 
The source-plane NIFS datacube is  re-binned adaptively by 5-10 to allow for \ha\ SNR $>$ 5 in each bin.  
(2): NIFS \ha\ intensity 2D map.  Note that the pixel scale of the source-plane  NIFS image is 0.$\!\!^{\prime\prime}$01.  The \ha\ image in panel (1)  is astrometrically aligned and re-binned to match the HST F814W source-plane resolution. 
(3): NIFS \ha\ velocity 2D map. 
(4): NIFS \ha\ velocity dispersion 2D  map.
\label{fig:fig4}}
\end{figure*}

\begin{figure*}[!ht]
\centerline{
\includegraphics[trim = 20mm 0mm 25mm 0mm, clip, width=0.99\textwidth,angle=0]{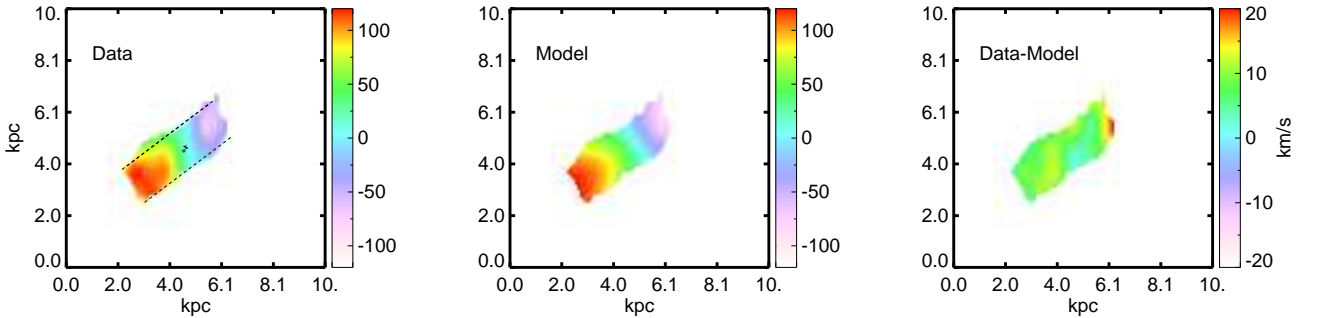}
}
\caption{ 2D disk model fit to the velocity field.  The left panel is the data. The best-fit  kinematic center is marked as a cross. The kinematic center matches the center defined 
by HST broad-band photometry  (shown as the star symbol) within  0.$\!\!^{\prime\prime}$1.  The dashed line displays a slit used to 
extract the 1D velocity in Figure~\ref{fig:fig6}.  Middle panel: best-fit model.  Right panel:  residual. 
\label{fig:fig5}}
\end{figure*}

\begin{figure}[!ht]
\centerline{
\includegraphics[trim = 5mm 10mm 8mm 8mm, clip, width=0.358\textwidth,angle=90]{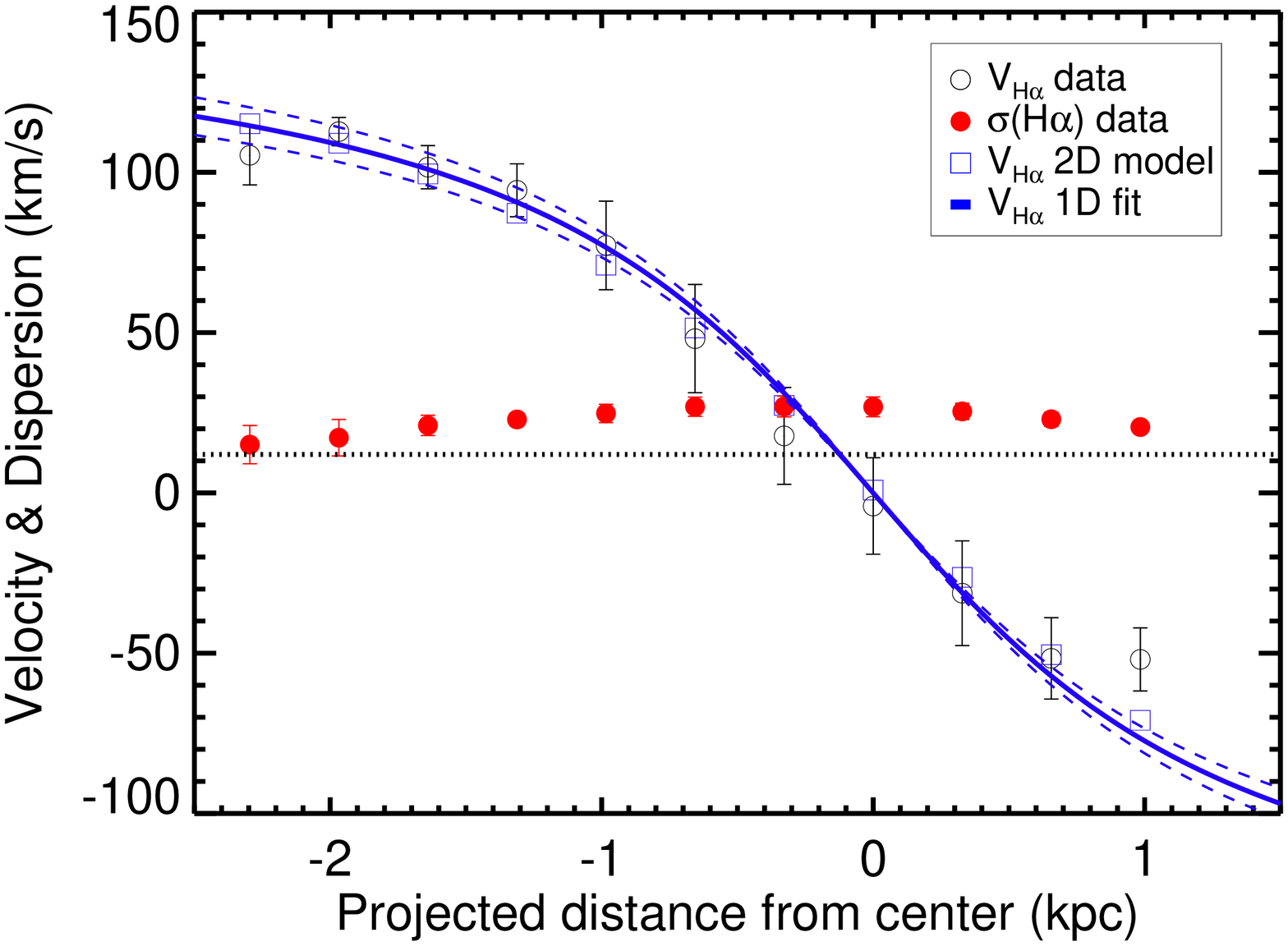}
}
\caption{One-dimensional \ha\ velocity (black empty circles) and velocity dispersion  (red filled circles) measured on a slit overlaid along the major axis (Figure~\ref{fig:fig5}). 
Each data point represents one resolution element along the major axis.  The error bars are standard deviations of spaxels binned perpendicular to the 
major axis.  The blue boxes are data extracted  from the 2D best-fit model. The best-fit 1D rotation curve and its 1$\sigma$ variation are shown as blue solid and dashed lines.  The horizontal black dotted line highlights the thermal
broadening threshold of the \ha\ line.  
\label{fig:fig6}}
\end{figure}

\section{Analysis and Results}\label{sec:results}
\subsection{NIFS Data Analysis}
\subsubsection{NIFS \ha\ Map and Kinematics} \label{subsec:map}
In Figure~\ref{fig:fig3} we show the 2D \ha\ flux intensity map  and 2D \ha\ kinematic maps on the observed frame (image plane), i.e., before lensing reconstruction.
To facilitate the comparison with HST images, we  mark the extent of  HST broad-band detections in contours in all panels.  We use the HST/F814W band image as an example because it has the deepest exposure among all HST filter observations.  We include the morphology of other HST bands in Section~\ref{subsec:hstmor}. 

In order to preserve the  spatial resolution offered by lensing and   because we have sufficient detections on individual spaxels, we  do not attempt to bin the original IFS datacube.   
The NIFS maps in Figure~\ref{fig:fig3}   are un-smoothed  and include only data with SNR $\ge$ 5.  \ha\  emission lines are detected  throughout the HST contours;  \ha\ emission lines are relatively  stronger near the galaxy core and star forming clumps in the spiral arm. 

The  rest-frame line-of-sight velocity 2D map is derived  from the \ha\ line centroids  with respect to the systematic redshift and the  velocity dispersion map from the \ha\ Gaussian line width. We use the median \ha\ line center to calculate the systematic redshift.  The instrumental profile has been subtracted in quadrature from the  best-fit Gaussian width to derive the intrinsic line width.   Figure~\ref{fig:fig3}  shows a systematic rotation and uniformly small velocity dispersion across the disk on the observed plane. 

The intrinsic morphology and NIFS measurements are shown in source-plane maps in Figure~\ref{fig:fig4}.
We use a grid subsampling of ${\it echant =20}$ on the image-plane and  ${\it s\_echant =10}$ on the source-plane in {\tt Lenstool} to 
 optimize the spatial resolution of the source-plane reconstruction.  The source-plane HST ACS image in  Figure~\ref{fig:fig4} is unbinned and has a pixel scale of  0.$\!\!^{\prime\prime}$005 after subsampling.  The source-plane NIFS datacube is rebinned adaptively by 5-11 pixels to allow for \ha\ line fitting of SNR $>$ 5 in each bin.   Note that because of the difference in the point spread function (PSF) on the image and source planes, the improvement in the source-plane SNR is usually less than what the magnification map predicts.  On average we 
 achieve a source-plane spatial resolution of $\sim$ 150 pc on the HST ACS images and $\sim$ 400 pc  on the NIFS 2D maps.

The source-plane 2D velocity map in  Figure~\ref{fig:fig4} clearly shows a velocity gradient consistent with a systematic rotation.   The velocity dispersion is enhanced slightly
in the kinematic center due to beam-smearing from the rotation. We estimate a maximum beam-smearing effect 
at  the kinematic center  to be  $\sim$ 24 \kms\  based on the velocity map of panel (3) in  Figure~\ref{fig:fig4}, consistent with the 2D velocity dispersion map of panel (4). 
    We do not see any significant  spatial correlation of the  \ha\ intensity map with the  velocity dispersion map.  Figure~\ref{fig:fig4} confirms that  {\it A1689B11} has a uniformly  low velocity dispersion across the disk on the source-plane. The mean velocity dispersion averaged
over all spaxels is  $V_{\rm \sigma, mean}$ {=} 23 $\pm$ 4 \kms.  Excluding  the central spaxels  that  are  affected most by beam-smearing, the average velocity 
dispersion on the outer disks is   $V_{\rm \sigma, outer-disk}${=} 15 $\pm$ 2 \kms.

\subsubsection{Disk Model Fitting} \label{subsec:diskmodel}
We use an empirically motivated arctangent function from \citet{Courteau97} to model the 2D velocity field: 
\begin{equation}
V(R) = V_0 + \frac{2}{\pi} V_c \arctan{\frac{R-R_0}{R_t}};
\label{eq:disk}
\end{equation}

The   source-plane line-of-sight velocity $v_{s}(R)$ is related to the intrinsic velocity $V(R)$  by the inclination angle $i$:
\begin{equation}
v_{s}(R) = V(R) sin (i);
\label{eq:inclin}
\end{equation}

The  source-plane radius vector $R_{s}$ is related to the intrinsic radius vector $R$ 
by:
\begin{equation}
\vec{R}_{s} = \vec{R}~ cos(i)
\begin{bmatrix}
    cos(PA), & -sin(PA) \\
    sin(PA), & cos(PA)
\end{bmatrix},
\label{eq:pa}
\end{equation}
where $PA$ is the positional angle.  For the definition of $PA$ and $i$,  we use the same convention  as  GALFIT \citep{Peng10},
i.e., $PA = 0$ is to the north (up), and $PA=90$ to the east (left);  $i=0$ is face-on and $i=90$ is edge-on.

The  seven free  parameters of the 2D disk model are thus:  central velocity $V_0$, 
inclination $i$, position angle $PA$, disk dynamic center $\vec{R_{0}}$
($R_{0x}$, $R_{0y}$), turn-over radius $R_t$, and 
asymptotic velocity $V_c$.    Note that $V_0$ is close to zero if the median \ha\ line center is a good approximation for the central velocity.
The best-fit model is obtained using  a $\chi^2$ minimization procedure similar to that described in \citet{Jones10a}. We use the statistical 1$\sigma$ errors
from the emission line fitting to compute the $\chi^2$.  The uncertainties of the best-fit parameters are estimated by perturbing the model until
the $\chi^2$ increases by one standard deviation from the best-fit model.

Figure~\ref{fig:fig5} shows the best-fit 2D disk model and the residual maps.  The reduced  $\chi^2$ of our best-fit model is 2.1 and the best-fit parameters are: 
asymptotic velocity $V_c$ {=} 200 $\pm$ 12 \kms, turn-over radius  $R_t$ {=} 1.7 $\pm$ 0.1 kpc, inclination angle $i=51 \pm 2$ degrees,
and positional angle $PA =  -37 \pm 2$ degrees.  The kinematic center is marked on Figure~\ref{fig:fig5} and is consistent with the brightest 
intensity from both the \ha\ image and the  HST broad-band images within 0.$\!\!^{\prime\prime}$1.

We can  also use Eq.~(\ref{eq:disk})  to fit the 1D  rotation curve.   We extract the 1D rotation velocities using a 0.$\!\!^{\prime\prime}$8 slit along the major axis with
the PA and the kinematic center  from the best-fit 2D model (see Figure~\ref{fig:fig5}, left panel).  The seven-parameter 2D model  then reduces to  a five-parameter 1D model.  In Figure~\ref{fig:fig6}, each data point represents one resolution element (0.$\!\!^{\prime\prime}$1) binned along the slit and  the error bars are the standard deviation of each data bin.    
 We show the best-fit 1D rotation curve and its 1$\sigma$ variations in Figure~\ref{fig:fig6}.  We find that the best-fit parameters from the 1D model are consistent with the  2D disk model within statistical errors. 
Note that we have also tested fitting a 3D disk model such as \citet{Teodoro15} to the datacube. Unfortunately,  we do not have sufficient SNR per spaxel for a reliable fit of the 3D model.  Our 2D and 1D disk model provides a reasonably good fit to the data.  Overall, a regular rotating disk  model is sufficient to explain
the velocity field of {\it A1689B11}.   The small residuals  on the edge of the disk  (Figure~\ref{fig:fig5} and Figure~\ref{fig:fig6}) are most likely due to under-estimation of the
observational noise.

The simple 2D disk model  does not consider beam-smearing and variations of the velocity dispersion.   However, as discussed in Section~\ref{subsec:map}, the disk shows a uniformly low velocity dispersion and the beam-smearing does not contribute significantly to the shape of the velocity field.    We show the 1D velocity dispersions (red points) along the kinematic  major axis  in  Figure~\ref{fig:fig6}.   The  velocity dispersion in the galactic center is mildly enhanced  due to beam-smearing effects.  The 
dispersion on the outer disk is approaching the thermal broadening threshold (12 \kms) of \ha\ emission line  (black dotted line).   

We discuss the kinematics of  {\it A1689B11} and compare it with other samples in Section~\ref{sec:disc}.

\begin{figure*}[!ht]
\centerline{
\includegraphics[trim = 5mm -5mm 5mm 5mm, clip, width=0.75\textwidth,angle=0]{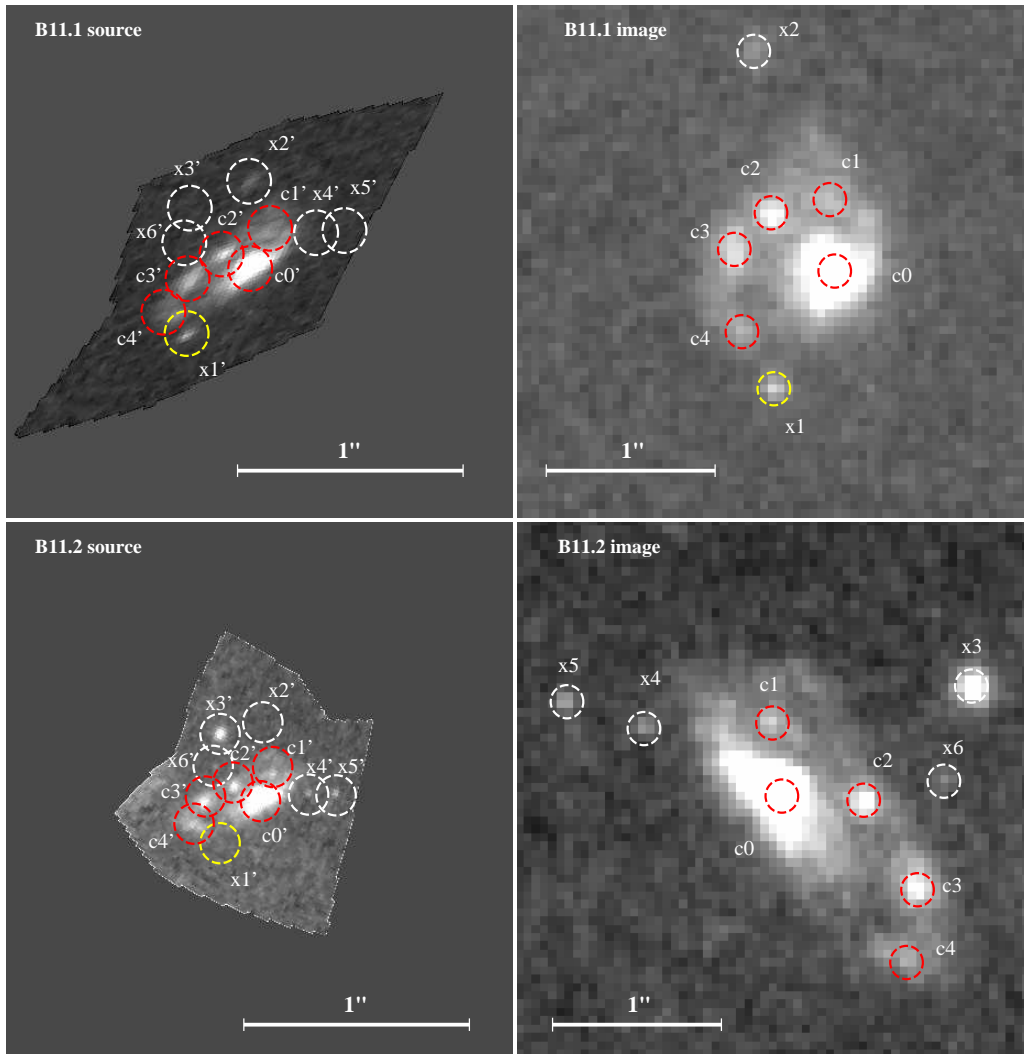}
}
\caption{Rejecting substructures that are not associated with the spiral disk by cross-comparing substructures on the image and source plane of   B11.1 and B11.2. 
Top left panel: source-plane morphology of B11.1.  Top right: image-plane morphology of B11.1.  Bottom left:   source-plane morphology of B11.2. Bottom right:
 image-plane morphology of B11.2. All images are based on the HST/F814W band.  The  circles on the right panels are manually identified clumps on the image-plane.
Corresponding positions of these clumps on the source-plane are predicted by lens models.  Clumps that are detected (SNR $>$ 5) in both of the left panels are bona fide substructures  (red circles) of the spiral galaxy, and vice-versa (white circles).  The yellow circle shows a clump that has a marginal (SNR $\sim$ 3) detection in B11.2 but is rejected because of the inconsistency in brightness.   All circles have a diameter of 0.$\!\!^{\prime\prime}$2, representing the RMS of our lens model reconstructions. 
\label{fig:fig7}}
\end{figure*}

\subsection{HST morphologies}\label{subsec:hstmor}
To compare our NIFS data with HST images in more detail, we analyze  source-plane morphologies for  all HST ACS and WFC3  bands.   B11.1 and B11.2 are well-detected 
in all four IRAC bands from Spitzer.  We do not gain resolved morphological
 knowledge from IRAC because of its poor spatial resolution ($\sim$ 2.$\!\!^{\prime\prime}$5, larger than the lensed images).  Therefore IRAC data is only used when deriving  global properties of {\it A1689B11} in Section~\ref{subsec:sfr}. 

We take full advantages of the multiple images from gravitational lensing and combine information from both images B11.1 and B11.2 to derive the best source-plane morphology.  B11.1 is covered by 6 filter bands (F475W, F625W, F775W, F814W, F105W, F140W) and  B11.2 is covered by these bands plus the additional  2 bands of F125W and F160W.  Source-plane morphologies for individual HST bands are reconstructed using the same method as implemented for the F814W  image described in Section 3.1.  
The spiral features are more prominent in the HST/ACS optical images (0.$\!\!^{\prime\prime}$05 resolution), whereas  HST/WFC3 images (0.$\!\!^{\prime\prime}$1 resolution) show mostly the central disk.  We include  source-plane morphologies for individual bands in the Appendix.

We  perform GALFIT \citep{Peng10} on the source-plane images  using a single exponential disk model. 
The fitting yields consistent morphological parameters  for all bands of HST images and for both B11.1 and B11.2.
The mean and standard deviation of the scale lengths ($r_{s}$) calculated from all 14 source-plane images are $r_{s}=1.3 \pm 0.4$ kpc, with an inclination angle of $i=55\pm10$ degrees and $PA =  -36 \pm 6$ degrees.   The single exponential disk model from GALFIT are in excellent agreement with geometric parameters derived from the NIFS  kinematic 2D disk model  (summarized in Table~\ref{tab:tab1}).  For the convenience of comparing with various definitions of radius in literature, we convert $r_{s}$ to the half-light radius $R_{1/2}$ and  the effective radius $R_{e}$ using the empirical relation of  $R_{1/2} = 2.2 r_{s}, R_{e} = 1.68 r_{s}$ \citep{Glazebrook13r}.   The half-light radius of {\it A1689B11} is therefore $2.6\pm0.7$ kpc.  
The central area of  {\it A1689B11} shows significant \ha\ emission from our NIFS data and some elongated substructures in the HST ACS images (Figure~\ref{fig:fig4}; Appendix). The central area could be a superposition of  a  star cluster and a bulge/bar component.    We  experimented with adding a bulge component to the exponential disk, however the fitting does not converge for most of the images.  We are currently investigating a more sophisticated  procedure for the bulge-disk decomposition and our preliminary result shows a very small bulge component.   We will report the full analysis in a future work focusing on the pitch-angle and bulge correlation of this spiral galaxy (Yuan, in preparation).

Because  gravitationally lensed images cover an extended area,
 the probability of having  contaminated foreground/background sources in the field of view of lensed images is larger
 than a non-lensed high-$z$ galaxy case. 
We  cross-compare substructures on the source-plane images of  B11.1 and B11.2 
to reject clumps/knots that may not be  associated with the spiral galaxy.   
We use the F814W HST images that have the deepest exposure and best spatial resolution for the cross-examination.   Figure~\ref{fig:fig7}
demonstrates our clump identification and rejection procedure. We first manually identify bright clumps on the right panels of  Figure~\ref{fig:fig7}.
There are 7 clumps (c0-c4, x1, x2) marked for lensed image B11.1 and 9 clumps (c0-c4, x3, x4,x5, x6) for lensed image B11.2.
 We then predict the source-plane positions and brightnesses of the clumps (c0'-c4', x1'-x6') and cross-examine them in the left panels of Figure~\ref{fig:fig7}.
 Because lensing conserves surface brightness and images B11.1 and B11.2 cover similar intrinsic areas of the lensed galaxy, the resolved source-plane substructures of B11.1 and B11.2 should be identical from the same HST observation \footnote{This method is not suitable in cases where the lensed images are crossed by critical lines, in which case the multiple lensed images may represent different parts of the intrinsic galaxy.}.   Source-plane clump brightness and  positions that are not consistent within the uncertainties of lens models are considered as interlopers.  For example,   clump  x2'  is identified  initially on B11.1 but is not detected on B11.2 within the lensing position uncertainty (RMS=0.$\!\!^{\prime\prime}$2) , we therefore reject clump x2' and consider it as a foreground or background source.    Similarly, clumps x4' and x5' identified initially on B11.2
 are not detected on B11.1 and are  rejected.    For clump x1' (yellow circles in  Figure~\ref{fig:fig7})  that is initially identified on B11.1, 
 there is a marginal 3$\sigma$ detection in the source-plane of B11.2, however,  the lens model predicts a source-plane brightness that 
 is $\sim$ 3 times brighter than what is observed.  We therefore reject x1' because of the inconsistency in flux magnification uncertainties ($\sim$ 10-20\%).   Note that clumps/knots  that are identified as foreground/background objects in this lensing analysis are at significantly different cosmological distances from the source galaxy and are therefore not satellites.   The final combined morphologies of B11.1 and B11.2  are shown in the Appendix.  
 
\begin{figure}[!ht]
\centerline{
\includegraphics[trim = 2mm 3mm 1mm 10mm, clip, width=0.57\textwidth,angle=0]{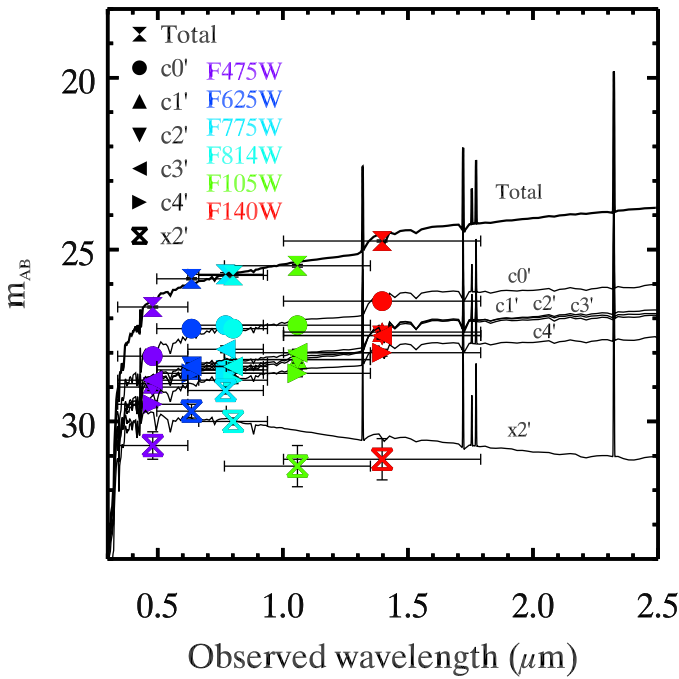}
}
\caption{SED fitting to the source-plane photometry for the entire galaxy (total) and individual clumps (c0'-c4',x2') identified in Figure~\ref{fig:fig7}.  
We show the six clumps that have reliable aperture photometry measured on the source-plane.     Details of the SED fitting are described in Section 3.2.  Clump x2'  has a very different SED shape compared to the total galaxy and the rest of the clumps, consistent with our  lensing position analysis that x2' is likely a foreground/background object.
\label{fig:fig8}}
\end{figure}

Finally, we  use  colors of the clumps as an alternative check for  substructure identities. 
We measure aperture photometry for all clumps in B11.1 in Figure~\ref{fig:fig7}  using  HST broadband  images.  
A reliable (SNR$>$5) photometry can  only be measured for six clumps across a minimal of 3 bands: c0'-c4' and x2'.    We  then carry out   Spectral Energy Distributions (SEDs) fitting for 
 the clumps  using the  software \verb+LE PHARE+ \citep{Ilbert09}. 
We fix the redshift at $z=2.54$ and use the stellar population synthesis models of \citet{BC03}.  We choose an initial mass function (IMF) of
\citet{Chabrier03} and the \citet{Calzetti00} attenuation law, with E$\rm{(B-V)}$ ranging from 0 to 2 and an exponentially decreasing star formation history.
The best-fit SEDs are shown in Figure~\ref{fig:fig8}.  Clumps c0'-c4' show similar SEDs, whereas x2' has a very different SED, consistent with our lensing source-plane position  analysis that
x2' is most likely an interloper.     We also show in Figure~\ref{fig:fig8} the best-fit SED for the total photometry of B11.1 (Section~\ref{subsec:sfr}).  
 In total,  we confirm 5 bona fide clumps (c0'-c4') for  {\it A1689B11}.  
There are significant \ha\ emissions detected on the central clump c0'  and on clump c2' on the spiral arm; \ha\ lines are also detected on clump c1' and c3' on the spiral arm (see Figure~\ref{fig:fig4}).  
   A  detailed analysis on individual clump properties will be reported in a separate work when we combine our ongoing OSIRIS/Keck data analysis on image B11.2. 
Our preliminary result shows that the  sizes ($400-600$ pc) and  surface star forming densities ($0.2-0.3$ $M_{\odot}$ yr$^{-1}$ kpc$^{-2}$)  of the clumps are comparable 
to high-redshift SF galaxies and are  in the intermediate range of what has been reported for     $z>1$ clumpy galaxies \citep[e.g.,][]{Genzel11,Jones10a,Livermore15}.  

\subsection{Star formation Rate, Stellar and Dynamic Mass}\label{subsec:sfr}
We estimate the total stellar mass from the SED fitting of broadband photometries from the HST ACS and  the Spitzer IRAC data \citep[details in][]{Yuan13a} (best-fit SED shown in 
Figure~\ref{fig:fig8}).  The best-fit stellar mass for {\it A1689B11} is \ms\  {=}  10$^{9.8\pm0.3}$$M_{\odot}$ and the best-fit extinction value is E(B-V)$_{\rm stellar}$ {=} 0.22. The total dust-corrected SFR from the SED fitting is SFR$_{\rm SED}$ {=} 22 $\pm$ 3 $M_{\odot}$ yr$^{-1}$.  All values have been corrected for  the lensing flux magnification.      The dust-uncorrected SFR from the total \ha\ fluxes of our NIFS observations is SFR$_{\rm H\alpha~no dust}$ {=} 3.9 $\pm$ 0.4 $M_{\odot}$ yr$^{-1}$.  Using the nebular dust extinction $E(B-V)_{\rm nebular}$ {=} 0.73 from
 \citep{Yuan13a} and the nebular attenuation curve of \citet{Cardelli89},   the  dust-corrected SFR from \ha\  is SFR$_{\rm H\alpha}$ {=} 22 $\pm$ 2 $M_{\odot}$ yr$^{-1}$,   in  agreement with  SFR$_{\rm SED}$. 
The total stellar mass and SFR of {\it A1689B11} are consistent with a  SF  galaxy on the $z\sim2$ mass-SFR relation (the main-sequence) within the $\sim$ 0.3 dex 1$\sigma$ scatter \citep[e.g.,][]{Zahid12b} of the relation.   The total SFR of {\it A1689B11}  is   10-20 times higher than a typical spiral galaxy of similar masses at $z\sim0$. 

Because we do not have spatially resolved dust attenuation measurements,  the  following estimation of SFR surface density is indirect and based on
a few assumptions.   Assuming the spatially resolved  $E(B-V)_{\rm nebular}$ is similar
to  the global $E(B-V)_{\rm nebular}$ measured from the slit data of \citet{Yuan13a}, then   the average SFR surface density is
$\Sigma$$_{\rm SFR}$ {=} 0.3 $M_{\odot}$ yr$^{-1}$ kpc$^{-2}$.    This value of  $\Sigma$$_{\rm SFR}$ is in the intermediate range of $z>1$ SF galaxies and is 1-2 orders of magnitude
higher than local SF galaxies \citep[e.g.,][]{Swinbank12,Fisher17,Zhou17}.
Assuming a simple Schmidt-Kennicutt (KS) relation \citep{Kennicutt98}, we convert  $\Sigma$$_{\rm SFR}$  into a gas surface density of $\Sigma$$_{\rm gas}$ $\sim$ 158 $M_{\odot}$  pc$^{-2}$. 
Assuming the surface area of the gas is $2\pi R^2$, where R is the radius where \ha\ are detected ($\sim$ 1.7 kpc), we then derive a 
 gas fraction ($f_{\rm gas}$ {=}  $M_{\rm gas}$/($M_{\rm gas}$+$M_{\rm star}$)) of   $\sim$ 18\%.   
 The  $f_{\rm gas}$ is in the lower range of star-forming galaxies at $z>1$ but still significantly higher than local SF galaxies \citep[e.g.,][]{Tacconi13}.  
 We compute the Toomre $Q$-parameter   \citep{Toomre64} for a gas-dominated disk as defined by 
  $Q \approx \kappa V_{\sigma}/\pi G \Sigma_{\rm gas}$, where $\kappa$ is the epicyclic frequency of the galaxy's rotation, $V_{\sigma}$ is the gas velocity dispersion, and $\Sigma_{\rm gas}$ is the surface mass density.  To compute $\kappa$, we assume a Keplerian disk so that $\kappa = \Omega$, where $\Omega$ is the angular frequency.
The $\Omega$ calculated at the half-light radius is $\Omega \sim 2 \pi/ 100 (Myr)$.   Adopting $V_{\sigma} = 23$ \kms, we  
  find $Q$$_{gas}$ $\sim$ 0.7 for {\it A1689B11}.   The small $Q_{gas}$  is commonly measured in  clumpy high-redshift SF galaxies and 
 is  consistent with the scenario that large SF clumps  form in the violent disk  instability  
 \citep[]{Genzel11,Law12n, Glazebrook13r, Shibuya16}.  We caution that the intrinsic $\Sigma$$_{\rm SFR}$ can easily change by a factor of two because of the unknown dust attenuation.    The systematic errors related to the methodology of deriving $f_{\rm gas}$ and $Q_{gas}$ are also highly uncertain and proper calculation requires estimation of the Mach number and molecular gas observations \citep[e.g.,][]{Federrath17}.

The dynamical mass assuming a rotationally supported disk is $M_{\rm dyn}${=} $RV$$^{2}$$_{\rm rot}$/$G$ {=}10$^{10.2 \pm 0.1}$$M_{\odot}$,  where $V_{\rm rot}$ is the 
asymptotic velocity $V_c$ and the  radius  $R$ is the turn-over radius $R_{t}$ from  the 2D disk model (Section~\ref{subsec:diskmodel}).   The ratio of the dynamical mass to the stellar mass for the inner $\sim$ 2 kpc  is therefore $M_{\rm dyn}/M_{\rm star} \sim 2.8$.  Taken the $M_{\rm dyn}$ as the sum of dark matter and baryonic matter, then the inferred dark matter mass fraction
within the inner $\sim$ 2 kpc (compatible to $R_{1/2}$) is $f_{DM} = 60\%$.  The value of $f_{DM}$ is in the typical range of local late-type spiral galaxies of
 similar $V_c$ and much 
larger than the massive baryon-dominated clumpy disk galaxies at $z\sim$ 2 \citep{Genzel17, Lang17}.  However,  this conclusion is subject to uncertain observational and  methodological errors of  $f_{DM}$. For example, by simply propagating the errors of  $M_{\rm dyn}$ and $M_{\rm star}$, we obtain  $f_{DM} = 60\% \pm 40\%$.  In addition, in order 
to decompose the contributions of the baryonic disk and the dark matter halo to the total rotation curve,  assumptions on the mass-to-light ratio (M/L) and scale-height of
 the stellar disk have to be made \citep[e.g,][]{Aniyan16}, both are difficult to constrain for our galaxy.  We therefore caution again that values derived in these two paragraphs are
dominated by systematic errors and should be interpreted with some caution.


\section{Discussion}\label{sec:disc}

\subsection{Spiral arms vs merger }\label{subsec:merger}
The morphology of {\it A1689B11} is indicative of a prototype spiral arm but not conclusive.  The clumpy morphology is also suggestive of a 
merger.  However, by combining the morphology with the high-spatial resolution kinematics presented in this work, we rule out the merger scenario
for  {\it A1689B11}.

Distinguishing mergers from isolated disks is extremely tricky at high redshift.  Neither morphological nor kinematic classifications alone can unambiguously exclude 
mergers.  Morphological classifications of mergers rely on footprints of interactions such as bridges, tidal tails and double nuclei. These features 
can be easily missed at high redshift  because of surface brightness dimming, size evolution and band shifting \citep[e.g.,][]{Hibbard97, Hung15}.  
High-redshift galaxies are clumpier and show more irregular structures than local galaxies, further complicating the visual
characterization of mergers \citep{ElmegreenD07}.    Kinematic classifications assume that isolated disks exhibit smooth velocity gradients whereas mergers show more asymmetric and chaotic  kinematic features \citep{Shapiro08,Dicaire08,Colina05}.    However, this assumption does not consider post-coalescence mergers which may also display disk-like kinematics \citep{Bellocchi12}.   A  ``morpho-kinematic" classification that combines the morphological and kinematic criteria is proposed as a more robust approach \citep{Rodrigues17}.

The velocity field of  {\it A1689B11} is consistent with a rotating isolated disk based on current  morphological and kinematic classification schemes.  We 
first use the kinematic criteria of the SINS survey \citep{Shapiro08} and  derive $V_{asym}$ and $\sigma_{asym}$  from the 2D velocity and velocity dispersion map.   We find $V_{asym} =0.1$ and $\sigma_{asym}=-0.08$,    placing {\it A1689B11} in the isolated disk region of the $V_{asym}$ versus $\sigma_{asym}$ diagram. 
The unique clump identification method presented in Section~\ref{subsec:hstmor} rejects most of the minor-merger like features as either foreground or background sources.  
For  clumps that do associate with  {\it A1689B11}, their velocities are consistent with rotating along with the main disk and their velocity dispersions
do not show  deviations from the rest of the disk, inconsistent with the scenario that the clumps are mergers.   To further rule out the scenario of post-coalescence mergers, 
we apply the five ``morpho-kinematic"   criteria of  \citet{Rodrigues17}.  We find  that {\it A1689B11} satisfies all five criteria of an isolated rotating disc:  (1) the velocity map
has a single velocity gradient (Section~\ref{subsec:diskmodel} and  Figure~\ref{fig:fig5}); (2) $V_{rot}/V_{\sigma} >1 $ ($=9-13$ for  {\it A1689B11});  (3) there is a $V_{\sigma}$-peak coinciding with the centre of rotation  (Section~\ref{subsec:diskmodel} and Figure~\ref{fig:fig4}); (4) there is no mismatch between the kinematic and morphological PAs, i.e., $\Delta PA < 30 \degree$ ($\Delta PA = 1 \degree$ for {\it A1689B11}, Section~\ref{subsec:map} and~\ref{subsec:hstmor}; Table~\ref{tab:tab1}); and (5) The rotation center matches the stellar mass center within 0.$\!\!^{\prime\prime}$4 ($\lesssim$0.$\!\!^{\prime\prime}$1 for {\it A1689B11}; Section~\ref{subsec:map} and~\ref{subsec:hstmor}; Figure~\ref{fig:fig5}).    Based on these arguments,
we therefore exclude  mergers as the origin of {\it A1689B11}.

\begin{figure*}[!ht]
\centerline{
\includegraphics[trim = 0mm 0mm 0mm 3mm, clip, width=0.75\textwidth,angle=0]{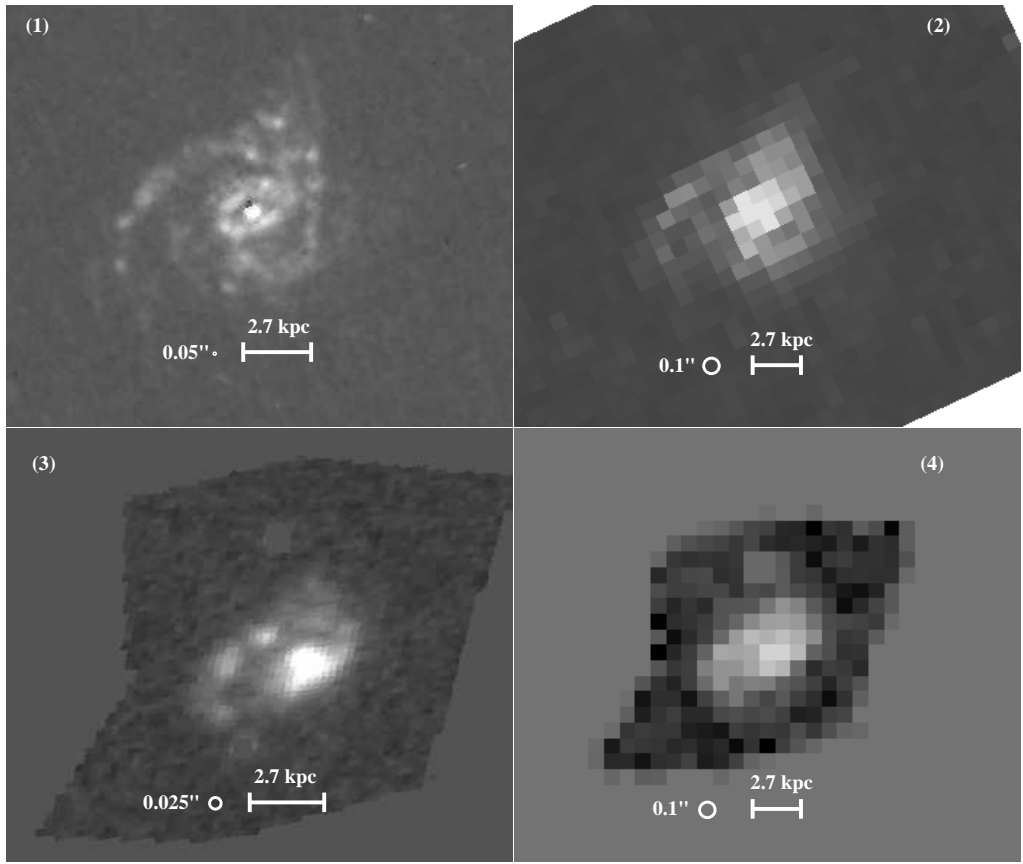}
}
\caption{An illustration of the effect of large SF clumps, surface brightness dimming and gravitational lensing.   (1)The HST/ACS narrow-band (\ha) image of a local spiral galaxy ``$G04-1$" that has large SF clumps similar to a $z\sim2$ clumpy disks (from the DYNAMO sample).  ``$G04-1$"  has a half-light radius of $2.7$ kpc, similar to {\it A1689B11}.   (2) The mocked HST/WFC3 IR image of  ``$G04-1$"  after being redshifted to $z=2.54$ without the effect of lensing. (3) The lens reconstructed morphology of  {\it A1689B11}.   (4) The   morphology of  {\it A1689B11} as it would 
appear in HST/WFC3 IR band without the lensing magnification.  The angular resolution and half-light radius are marked on each panel. 
\label{fig:fig9}}
\end{figure*}

\begin{figure}[!ht]
\centerline{
\includegraphics[trim = 10mm 10mm 10mm 20mm, clip, width=0.42\textwidth,angle=90]{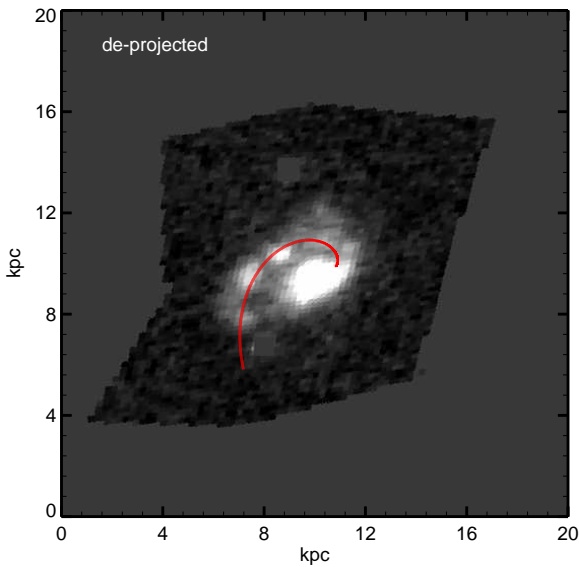}
}
\caption{The best-fit logarithmic spiral function (red line) to the de-projected image of {\it A1689B11}.  The 
best-fit pitch angle is $\theta = 37\degree \pm 2\degree$.   
\label{fig:fig10}}
\end{figure}

\subsection{Spiral arms vs clumpy disks and irregulars}\label{subsec:clump}
Spiral galaxies and irregular galaxies are two distinct morphological classes in the local universe \citep[e.g.,][]{Hubble26}.  However, as the general morphology of distant galaxies becomes more  chaotic and irregular \citep[e.g.,][]{Abraham96,Abraham96b,Conselice05,ElmegreenD07,Shibuya16}, 
spirals and clumpy/irregular galaxies do not have to be mutually exclusive at high redshift.  For example,  studies focused on spiral morphologies show that spiral structures are highly disturbed  and arms  are less well-developed at at $z \gtrsim 0.5$ \citep{Abraham01,ElmegreenD05}.  The less well-defined spiral structures at high redshift could be caused by either   intrinsic evolutions and/or observational biases.  

Spiral arms are the main sites of star formation in the local universe.  
The brightness, sizes and  surface densities of star forming regions  are  much larger at high redshift  \citep[e.g.,][]{ElmegreenD05,Jones10a, Genzel11,Swinbank12,Fisher14}. 
The  size and brightness of the star forming clumps in the spiral arm   of  {\it A1689B11}  are comparable to those of  $z>1$ clumpy SF galaxies (Yuan, in preparation),  producing  a much clumpier  appearance than  local spiral arms.  In addition, 
{\it A1689B11} is gravitationally lensed; the reconstructed source-plane morphology is much sharper than a non-lensed case, making it look different from most $z\sim$ 2 SF galaxies.

To demonstrate the effect of large star forming clumps  and observational effects such as surface brightness dimming and gravitational lensing on the appearance of spiral arms at high redshift, we manually redshift a local spiral galaxy to $z=2.54$ and compare its redshifted morphology with the lensed and non-lensed case of  {\it A1689B11} (Figure~\ref{fig:fig9}).  
 The local spiral galaxy template ($G04-1$) is chosen from  the DYNAMO (DYnamics of Newly-Assembled
Massive Objects) sample.  The DYNAMO sample is a local analog of  turbulent, clumpy disk galaxies  at high redshift \citep{Green10,Green14,Bassett14}. 
  $G04-1$ is one of the few  galaxies in the DYNAMO sample that have a spiral morphology.  $G04-1$  has similar half-light radius ($R_{1/2,H\alpha} = 2.7$ kpc),  star forming clump size and brightness as {\it A1689B11} \citep{Fisher17}.  We use the HST \ha\ narrow-band  image (pixel  size $0.\!\!^{\prime\prime}05$) of $G04-1$ as it best  represents  the clumpy SF morphology (Figure~\ref{fig:fig9}, panel (1)).   
   Figure~\ref{fig:fig9}-panel (2) illustrates the mocked HST/WFC3 (pixel size $0.\!\!^{\prime\prime}1$) morphology of  $G04-1$ at $z=2.54$. We assume no intrinsic size evolution because the DYNAMO sample consists of compact objects in the local universe,  analogous to the sizes of galaxies $z\sim2$. 
Because  $G04-1$ is nearly face-on, the morphology  of  {\it A1689B11}  is deprojected in panels (3) and (4)  for a better comparison.  
Panel (3) of Figure~\ref{fig:fig9} shows  the best source-plane reconstructed morphology of {\it A1689B11} based on the HST/F814W band image. Owing to gravitational lensing,
the  effective spatial resolution on the source-plane is increased by a factor of 2-3 and the SNR of the image increased by $\sim$ 7.   To compare with the non-lensed image
of $G04-1$ at $z=2.54$, we show in panel (4) what {\it A1689B11}  would have looked like without lensing magnification in the HST/WFC3 IR band.

Panel (1) and (2) of Figure~\ref{fig:fig9} underline the importance of separating  observational  effects from intrinsic evolution of spiral arms. 
In the local universe, spiral galaxies are divided into three classes based on the number of arms:   grand-design (two-arm), many-arm (multiple distinct global arms), and 
flocculent (multiple less distinct arms) \citep[e.g.,][]{ElmegreenD82,Hart17}.  According to the local image of Figure~\ref{fig:fig9}-panel (1), $G04-1$ is classified as a  many-arm spiral,
as are more than half of the  local spiral galaxy population \citep{Davis14}.  As a result of surface brightness dimming,  $G04-1$ would be mostly likely classified as a  rare one-arm spiral that are only seen in $14\%$ of local spirals \citep{Davis14}.  It is  possible that  {\it A1689B11}  has more arms like $G04-1$, but  only the longest arm is visible at this redshift and with this magnification.

Panel (3) and (4) of Figure~\ref{fig:fig9} demonstrate  the power of gravitational lensing in bringing the otherwise unseen spiral structures into focus. 
 The spiral structure is still detectable in the redshifted image of $G04-1$, whereas  the spiral arm of  {\it A1689B11}  is  barely distinguishable in the un-lensed case of panel (4). 
 Note that although  $G04-1$ and {\it A1689B11} share similar galaxy size and SF clump brightness, $G04-1$  is  $\sim$ 10 times more massive than {\it A1689B11} and has a large \ha\ velocity dispersion of $\sim$ 50 \kms.  There is noticeable differences in the morphological appearance of panel (3) and (4), implying that the intrinsic spiral structures 
of  {\it A1689B11} are  less well-developed than $G04-1$.    While a detailed comparison of the host properties of  $G04-1$ and {\it A1689B11} is beyond of the focus of the current paper, 
 Figure~\ref{fig:fig9} simply demonstrates that the morphology  of   {\it A1689B11} is consistent with a spiral galaxy. The effect of gravitational lensing, surface brightness dimming and larger SF regions at high-redshift combined together to make the somewhat unique appearance of {\it A1689B11}.  
 
Finally, separating spiral arms  from clumpy irregular morphologies requires a quantitative classification scheme.
Such a scheme is not available yet at  high redshift.  For example, there are  two geometric parameters that are commonly
used in quantifying local spiral arms: the number of arms (or harmonic modes) and the pitch angle.  A large fraction of local spirals
can be modeled by superpositions of  logarithmic spiral functions.  
Based on this, automatic spiral arm finding and logarithmic function fitting tools have been developed and applied in
 local galaxy surveys \citep[e.g.,][]{Davis12,DavisD14,Shields15,Hart17}.    
Our on-going   effort of testing and adapting these tools at high-redshift will help to quantify spiral arms and to separate spiral from irregular structures objectively.
We show in Figure~\ref{fig:fig10} a preliminary best-fit logarithmic spiral function created by manually masking pixels that form the spiral arm. We 
  find a  large pitch  angle ($\theta = 37\degree \pm 2\degree$)  for the arm.   We test the automated arm detection and fitting routine of SpArcFiRe \citep{DavisD14} on {\it A1689B11} 
and notice that  the number of arms and pitch angle depend sensitively on the bulge-disk decomposition, lensing PSF reconstruction and masking of noise.   We will report the
pitch angle analysis in our future work (Yuan, in preparation).   To conclude, we favor the interpretation of a spiral galaxy with primitive spiral arms developing in {\it A1689B11} instead
of a merger or an irregular galaxy.

\begin{figure*}[!ht]
\centerline{
\includegraphics[trim = 0mm 12mm 1mm 0mm, clip, width=0.7\textwidth,angle=90]{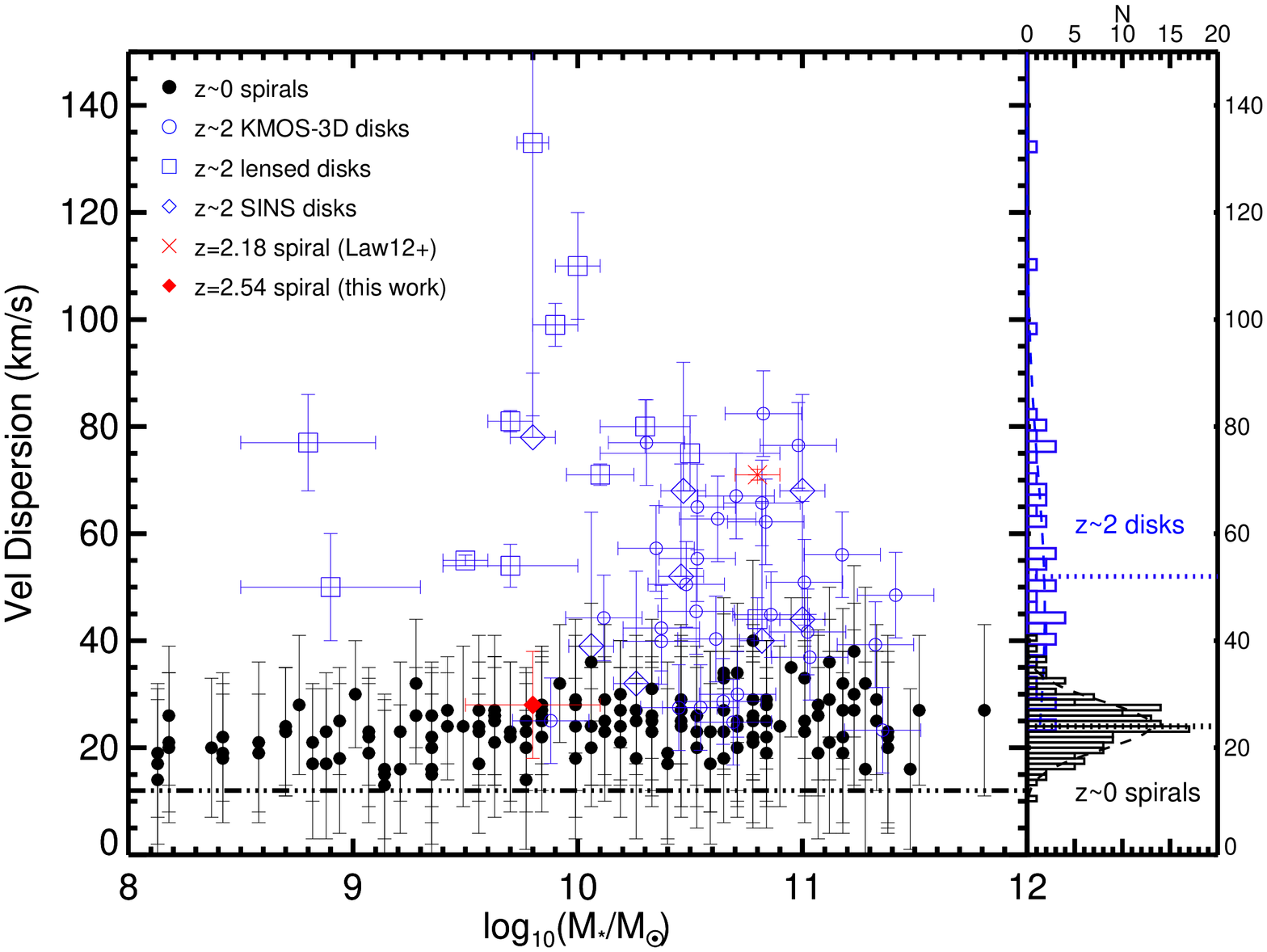}
}
\caption{ H$\alpha$ velocity dispersion versus stellar mass for  {\it A1689B11} (the red diamond) and comparison samples at  $z\sim0$ and $z>2$.  
The blue empty diamonds are the $z>2$ SINS disks \citep{FS06,FS09,Cresci09}; the blue empty circles are the $z>2$ KMOS-3D disks \citep{Wisnioski15},
the blue empty squares are  gravitationally lensed galaxies at $z>2$ \citep{Jones10a,Livermore15}.  The local spiral galaxy  sample (black filled circles) is from  \citet{Epinat10a}.
The red cross shows the only other known $z>2$ spiral, BX442, detailed in \citet{Law12n}.  {\it A1689B11} is marked by the red filled diamond. The horizontal  dash-dotted line 
indicates the minimal velocity dispersion that can probed by \ha\ lines because of the thermal broadening  ionized   gas ($10^{4}$ K).  The panel on the right shows the velocity dispersion distribution
of the $z\sim$0 and $z\sim$2 samples. 
\label{fig:fig11}}
\end{figure*}

\subsection{Comparison with other samples}\label{subsec:compare}
The total SFR (22 $M_{\odot}$ yr$^{-1}$) and stellar mass (10$^{9.8}$ $M_{\odot}$) place {\it A1689B11} as a typical SF main-sequence galaxy
at $z\sim2$.   However, the kinematic properties of   {\it A1689B11}  are quite ``mature" compared to a $z>1$ SF galaxies and are more akin to  local spiral galaxies.     

We compare the gas  velocity dispersion and stellar mass of  {\it A1689B11} with other samples in Figure~\ref{fig:fig11}.  We collect  \ha\  gas  velocity dispersion measurements on disk galaxies from a few representative IFS surveys at $z\gtrsim2$:  the SINS disks \citep{FS06,FS09,Cresci09}, the KMOS-3D disks \citep{Wisnioski15}, and gravitationally lensed galaxies \citep{Jones10a,Livermore15}. Only the rotation-dominated ($V_{rot}/V_{\sigma} >1$) disks from these surveys are shown.     The  $z\sim0$ sample is taken from a local reference sample of rotating spiral galaxies with \ha\ velocity dispersion measurements \citep{Epinat10a}.

The local spirals  have low \ha\ velocity dispersions irrespective of their stellar masses. 
There is a large spread in velocity dispersions of  $z\gtrsim2$  disks, with a median value that is $\sim$ 2.5 times larger than local spirals.  
We also mark the thermal broadening of \ha\ emitting  gas ($10^{4}$ K) as the minimal velocity dispersion that can be measured from the \ha\ line. 
The systematic uncertainty in comparing velocity dispersions across  samples could also contribute to the large scatter in high-redshift samples.  The major systematics  come from
the method used to correct for beam-smearing,  the disk model, and the radius at which the velocity dispersion is measured and the weighting used.  
The velocity dispersion of {\it A1689B11} is  considerably lower than  lensed galaxies of similar masses.
It is also much lower than the median of all $z\gtrsim2$  samples.   The large  $V_{rot}/V_{\sigma}$ ($\sim9-13$) of  {\it A1689B11}  is similar to local spirals and  $\sim$ 2-5 times larger than the median value of SF galaxies at $z\gtrsim1$ \citep[e.g.,][]{Cresci09,Swinbank17}.

One of the major observational results from IFS surveys of high-redshift ($1 \lesssim z \lesssim 3$) galaxies is that the intrinsic gas velocity dispersions as usually measured from optical ionized gas are significantly higher  than local SF galaxies \citep[e.g.,][]{Law09, Lehnert09, FS09, Genzel11,Wisnioski15}.    Popular explanations for the enhanced 
velocity dispersions include  star formation feedback \citep{Lehnert09, Green14},  gravitational instability \citep{Agertz09,Ceverino10}, a combination of star formation feedback driven and gravitationally driven turbulence \citep{Krumholz16,Krumholz17},    a multitude of physical drivers of turbulence  \citep{Federrath12, Federrath17b} 
 and cosmological cold gas accretion  \citep{Dekel09, Bournaud09, Genzel11, Genel12}.

If the ionized gas  turbulence of {\it A1689B11} is directly driven by energy ejected from star formation through supernova,  then 
its  SFR surface density suggests  a  velocity dispersion of $V_{\sigma} \sim 40-55$ \kms  \citep[e.g.,][]{Lehnert09,Green14}, depending on how   the energy is dissipated into the ISM and assuming a conservative  supernova feedback efficiency of 0.25 \citep{Dib06,Zhou17}.  In this simple model, the small  velocity dispersion of {\it A1689B11}  would imply that the supernova feedback efficiency is a factor of $\sim$ 2 lower than local galaxies, which is difficult to explain.  

In the slightly more complicated  feedback driven turbulence model of \citet{Krumholz16}, where 
$V_{\sigma}$ is a function of SFR and $Q_{gas}$, the small velocity dispersion of  {\it A1689B11} can be reproduced at a $Q_{gas} \sim 0.5$, matching our rough estimation of $Q_{gas}$ in Section~\ref{subsec:sfr}.  
The gravity driven  turbulence model of  \citet{Krumholz16} is disfavored because the small $V_{\sigma}$ would require an unrealistically large gas fraction ($50\% < f_{gas} \sim 100\%$).
Future direct observations of molecular gas and dust maps of {\it A1689B11} would provide a more robust measurement on $Q_{gas}$ and gas fraction to distinguish various turbulence driven models. 

 We also show in Figure~\ref{fig:fig11} the location of the spiral galaxy  {\it BX442} at $z {=} 2.17$ from \citet{Law12n}. 
  {\it BX442}  has a high velocity dispersion ($V_{\sigma}$ $\sim$ 70 \kms) and is thought to have a short-lived spiral  triggered by a minor merger \citep{Law12n}. 
  Two conditions must be satisfied according to  the minor-merger trigger mechanism.  First the galaxy must be massive enough to stabilize the formation of an extended disk. Second,  a nearby merging satellite  must be properly orientated and sufficiently massive  to excite the spiral feature.   {\it A1689B11} is $\sim$ 10 times less massive than {\it BX442} and is not consistent with a merging system as discussed in Section~\ref{subsec:merger}.   Even if the minor merger scenario works for {\it BX442}, an  alternative  mechanism  is required to account for the existence of  {\it A1689B11}.

\subsection{Spiral Arm and Thin Disk formation}\label{subsec:thindisk}

Being an outlier in both the velocity dispersion distribution and morphology (spiral structure) of high-redshift galaxies, {\it A1689B11}  
provides a few interesting angles to revisit the formation of spiral arms.

Classic theories of spiral arm formation require a dynamically cool and thin  disk.  In the paradigm of the density wave theory,   the amplitude of  the induced density wave  
 becomes too weak  if  the gas  velocity dispersion is too high or  if the disk is too thick \citep[e.g.,][]{Lin64,BertinS96,ElmegreenB93,Rafikov01,Bottema03,Sellwood14}. 
 In the alternative theory of swing amplification or self-gravitational instability, a large velocity dispersion (in the context of a large Toomre-$Q$ parameter) dampens the amplification and a
 thick disk reduces  the disturbance gravity that are seeds of the swing amplification  \citep[e.g.,][]{Julian66,ElmegreenB93,Bottema03,Sellwood14}.
In the local universe, almost all spiral arms reside in the thin disk (scale height 200-300 pc) with a dynamically cool stellar  ($V_{\sigma}$(star)$\sim20$ km/s) and gas component ($V_{\sigma}$(H$\alpha$)$\sim20-25$ km/s) \citep{Glazebrook13r}. For  stellar disks that are dynamically ``warm" and have a non-negligible thickness, spiral activity develops 
mostly in the cold gas component \citep{BertinS96}.  The exceptions are mergers or bar induced spirals, where the
 external driver plays a more important role than internal disk dynamics \citep[e.g.,][]{Kormendy79,Bottema03,Dobbs10}. 

Both the density wave  and swing amplification theory have seen  a certain degree of success in explaining local spirals where the model assumptions are easily satisfied. The  logarithmic spiral function predicted from the density wave theory has been widely used and confirmed in observations of local spirals, whereas swing amplification is  commonly used to explain transient and less regular spirals.     The main difference is that  density wave theory predicts quasi-steady long-lived arms (wave arm) and swing amplification or its various form of gravitational instability  predicts short-lived reoccurring arms (material arm).  It is difficult to  constrain the lifetime of the spiral arm observationally, thus hard to distinguish the two theories \citep[e.g.,][]{Sellwood11}.    Current simulations     seem to prefer either swing amplification or merger in explaining the formation of spiral arms at  high redshift \citep[e.g.,][]{Law12n,Fiacconi15}.     
For example,  self-gravitational  instability and fragmentation has been proposed to model  the local DYNAMO spiral  $G04-1$ (Section
~\ref{subsec:clump}) that has  a high velocity dispersion and clumpy SF regions analogous to high-redshift disks \citep{Inoue17}.  Minor merger is thought to trigger the short-lived spiral structure of  {\it BX442} \citep{Law12n}.

It is reasonable to expect that  the quasi-steady  spiral arm from the density wave theory is  suppressed at high redshift   
because of the high gas velocity dispersions and geometrically thick disks  \citep{Cresci09,ElmegreenB06}.   The rarity
of spirals at $z\gtrsim2$ could be partly accounted for by the short-lived arms from either gravitational  instability or mergers.  
On the other hand, the discovery  of a dynamically cool disk like {\it A1689B11} could mean that  the condition for the 
classic  density wave spirals to develop  can exist at $z>2$.  The small velocity dispersion of {\it A1689B11} is indicative of a thin disk (Figure~\ref{fig:fig11}). 
 Our preliminary fitting of a logarithmic spiral function (Figure~\ref{fig:fig10})  is consistent with
a density wave triggered primitive spiral arm of {\it A1689B11}.
We speculate that {\it A1689B11} belongs to a population of rare spiral galaxies at $z\gtrsim2$ that mark the earliest epoch of thin disk formation \citep{Freeman02,Kraljic12,Freeman12, Elmegreen17}. 
Note that the spiral structure of {\it A1689B11} would not be visible with current observational capacity without gravitational lensing (Figure~\ref{fig:fig9}). 
Future observations with the James Webb Space Telescope (JWST) will help to reveal this population and  investigate the earliest onset of spiral arms and  thin disks.

\subsection{The origin of  spiral arms in cosmological zoom-in simulations }\label{subsec:simu}
Modern theoretical efforts in understanding the origin of spiral arms have  focused on semi-analytic or N-body simulations in isolated disks \citep[e.g.,][]{Wada11,Baba13,DOnghia13,DOnghia15}.  Cosmological simulations do not have sufficient resolutions to trace the detail dynamics of spiral arms, however, 
it is interesting to explore the spiral arm formation from the context of cosmological simulations where the role of environment is  included. 

Recent  cosmological zoom-in simulations \citep{Cen14a} provide valuable insights into the emergence of the Hubble sequence across cosmic time.  
\citet{Cen14a} shows the rich physics of cold gas accretion dynamics onto galaxies and evolutionary trends.   
At $z\gtrsim2$,  the average in-situ  cold gas accretion streams
through the galactic halo virial sphere surface can be characterized  by:  (1) multiple cold streams, (2) high accretion rates, (3) low angular momenta, and (4) high 
gas densities.  Among these four parameters,   it is suggested that spiral structures are most sensitive to (1) the number of concurrent streams in the gas accretion.    
Flocculent spirals only begin to significantly appear at $z\sim 1-2$ when the number of major gas streams are about two to three
and two-arm grand-design spiral galaxies appear at $z\leq$ 1 when the number of major cold streams reduces to one.  
In the framework of \citet{Cen14a},  the average number of gas streams decreases from high to low redshift. 
However, at each redshift, there is a distribution of in-situ environments pertaining to the range of cold streams.
It is therefore possible to find ``evolved" spiral galaxies at high-$z$ and ``unevolved" high-redshift analog galaxies at low-$z$ at the tail of the environmental distributions.   

In the framework of \citet{Cen14a},  spiral {\it A1689B11} may be formed in an in-situ environment that is characterized by probably one major stream with high accretion rates and gas densities for an extended  period of time. Because the number of significant cold streams is correlated with the degree of interactions among galaxies, it is expected that the velocity dispersion of the stellar disk
would be positively correlated with the number of cold streams \citep{Cen14a}.  This scenario is consistent with accretion energy being one of the main drivers of the  turbulence in disks at $z\sim$2 \citep[e.g.,][]{Birnboim03, Dekel09,Bournaud09,Genel12}. 

However, galaxy growth through cold-mode accretion  has been challenged by other cosmological simulations \citep[e.g.,][]{Nelson13,Genel14}. 
Whether the Hubble sequence originates from the cold flow process or other mechanisms such as mergers and feedback is  highly controversial \citep[e.g.,][]{Genel15}. 
Moreover, extra care must be given  when studying   the origin of the spiral arms in cosmological  simulations. 
Spiral features in current cosmological simulations may originate from unphysical perturbations  and  are sensitive to the resolution and the  detailed prescriptions of ISM models. 
With a larger sample of high-redshift spirals in observations and larger volumes of  cosmological zoom-in simulations with sub-pc resolutions, we  should be able to answer the question
of whether the in-situ environment plays a critical role in the formation of spiral galaxies.

 \begin{table}
\begin{center}
\caption{Physical properties of {\it A1689B11}}
\label{tab:tab1}
\begin{tabular}{lc}
\hline
\hline
RA (J2000) & 13:11:33.336\\
DEC(J2000)      &  -01:21:06.9  \\
\hline
{\it Spectroscopic data measurements:} & \\
 \hline
Redshift (z$_{H\alpha}$)   & 2.540\\
Velocity dispersion ($V_{\rm \sigma}$, mean)   & 23$\pm$ 4 \kms \\
Velocity dispersion ($V_{\rm \sigma}$, outer disk 1-2 kpc)   & 15$\pm$ 2 \kms \\
Star formation rate (SFR$_{\rm H\alpha}$)  & 22 $\pm$ 2 $M_{\odot}$ yr$^{-1}$\\
Dust attenuation (E(B-V)$_{\rm Balmer~decrement}$)  & 0.73 \\
\hline
 {\it   GALFIT/exponential disk best-fit: } & \\
 \hline
Inclination  ($i$)  &  55 $\pm$10 degrees\\
Position angle  (PA)  &  -36 $\pm$6 degrees\\
Scale length ($r_{s}$) & 1.3$\pm$0.4  kpc\\
Half-light radius ($R_{1/2}$)  & 2.6$\pm$0.7  kpc\\
Effective radius ($R_{e}$)  & 2.0$\pm$0.4  kpc\\
\hline
 {\it 2D velocity disk model best-fit:}  & \\
 \hline
Inclination  ($i$,velocity)  &  51 $\pm$2 degrees\\
Position angle  (PA,velocity)  &  -37 $\pm$2 degrees\\
Radius rotation ($R_{t}$, velocity) & 1.7 $\pm$ 0.1 kpc\\
Rotation Velocity ($V_{rot}$)  & 200 $\pm$ 12 \kms \\
\hline
 {\it Inferred physical parameters:} & \\
 \hline
Star formation rate (SFR$_{\rm SED}$)  &  22 $\pm$ 3 $M_{\odot}$ yr$^{-1}$\\
Stellar mass ($M_{\rm star}$) & 10$^{9.8\pm0.3}$ $M_{\odot}$ \\
Dynamical mass ($M_{\rm dyn}$)  & 10$^{10.2}$ $M_{\odot}$\\
 Average SFR surface density ($\Sigma$$_{\rm SFR}$)  & 0.3 $M_{\odot}$ yr$^{-1}$ ~kpc$^{-2}$ \\
 Average gas surface density ($\Sigma$$_{\rm gas}$)  & 158 $M_{\odot}$  ~pc$^{-2}$ \\
 Gas fraction ($f_{\rm gas}$)   & $\sim$ 0.18 \\
Toomre parameter Q$_{\rm gas}$ & $\sim$ 0.7 \\  
\hline
\hline
\end{tabular}
\end{center}
\small
{\sc Notes. ---}  
All values  have been corrected for  lensing  magnifications.  PA are defined as: PA{=}0 when the major axis   is up and positive
if rotated counterclockwise.
Assumptions for  inferred values are described in Section 3.
\vspace{0.1cm}
\end{table}

\section{Summary and Future Work}\label{sec:future}
We report  NIFS/Gemini observations on a $z=2.54$ gravitationally lensed spiral galaxy {\it A1689B11}.
It is the highest redshift spiral galaxy observed with the highest spatial resolution  and spectroscopic depth  to date. 
{\it A1689B11} shows a primitive spiral arm that is scarcely seen in other galaxies at $z\gtrsim2$.
Regarding the SFR, size and stellar mass,   {\it A1689B11} is representative of a $z\sim$ 2 SF galaxy. 
In contrast,  the \ha\ kinematic field shows striking similarities to $z\sim0$ isolated late-type spiral galaxies. 
It shows an ordered rotation  ($V_{\rm c}$ {=} 200 $\pm$ 12 \kms) and  uniformly small velocity dispersions ($V_{\rm \sigma, mean}$ {=} 23 $\pm$ 4 \kms and $V_{\rm \sigma, outer-disk}$ {=} 15 $\pm$ 2 \kms).     The low gas velocity dispersion is consistent with  the classic density wave theory that  spiral arms form in dynamically cold and thin disks. 
    We speculate that {\it A1689B11} belongs to a population of rare spiral galaxies at $z\gtrsim2$ that mark the formation epoch of thin disks. Future observations with JWST will help to reveal this population and    investigate the earliest onset of spiral arms.

Our follow-up work include a detailed study on the angular momentum and SF clump properties by including  our recent  OSIRIS observation on the more magnified image B11.2 of {\it A1689B11}.  We are exploring a more robust method of bulge/disk decomposition of {\it A1689B11} and  plan to investigate the bulge to pitch angle correlation in a follow-up paper (Yuan, in prep).  Our on-going IFS observations on a larger sample of non-lensed spiral galaxies ($\sim$ 30) at  $z\gtrsim2$ that we recently discovered  will help to validate/reject our speculations about the origin and number density of spiral galaxies like {\it A1689B11}.

\acknowledgments 
We thank the anonymous referee for his/her report, which helped to restructure and improve the work significantly.
This work is a tribute to the late Peter McGregor who built NIFS/Gemini and taught TY how to reduce the NIFS data.  
This research was conducted by the Australian Research Council Centre of Excellence for All Sky Astrophysics in 3 Dimensions (ASTRO 3D), through project number CE170100013.
TY thanks useful discussions with  Ken Freeman, Lars Hernquist, Enrico Teodoro, Yusuke Fujimoto, Ben Davis and the GEARS3D group.   
TY acknowledges the support from the ASTRO 3D fellowship. 
BG gratefully acknowledges the support of the Australian Research Council as the recipient of a Future Fellowship (FT140101202).
CF acknowledges funding provided by the ARC Discovery Projects (grants~DP150104329 and~DP170100603).
JR acknowledges support from the ERC starting grant 336736-CALENDS.
LK gratefully acknowledges support from an Australian Research Council (ARC) Laureate Fellowship (FL150100113).
RC acknowledges grants NNX12AF91G and AST15-15389. 
YB acknowledges ISF grant 1059/14. 
We thank Stuart Ryder for his assistance with the Gemini observations.

The data for this work  is based on observations obtained at the Gemini Observatory, which is operated by the Association of Universities for Research in Astronomy, Inc., under a cooperative agreement with the NSF on behalf of the Gemini partnership: the National Science Foundation (United States), the National Research Council (Canada), CONICYT (Chile), Ministerio de Ciencia, Tecnolog\'{i}a e Innovaci\'{o}n Productiva (Argentina), and Minist\'{e}rio da Ci\^{e}ncia, Tecnologia e Inova\c{c}\~{a}o (Brazil). The authors wish to recognize and acknowledge the very significant cultural role and reverence that the
summit of Mauna Kea has always had within the indigenous Hawaiian
community. 

{\it Facilities:} \facility{Gemini  (NIFS)}.

\begin{figure*}[!ht]
\centerline{
\includegraphics[trim = -10mm 0mm 1mm 0mm, clip, width=0.75\textwidth,angle=0]{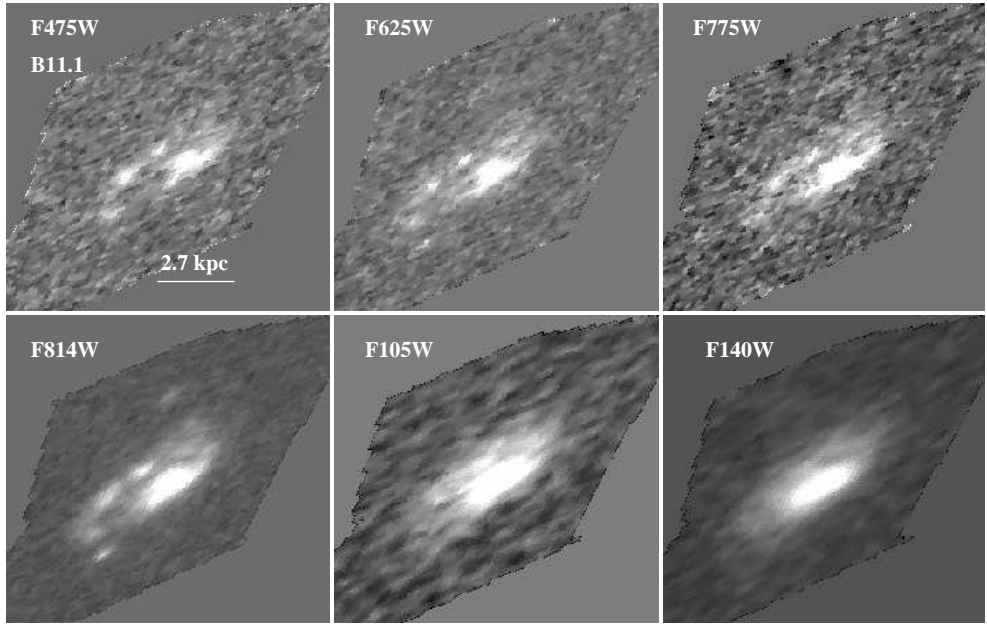}
}
\caption{ Multi-band HST images of B11.1 on the source-plane. 
\label{fig:figa1}}
\end{figure*}

\begin{figure*}[!ht]
\centerline{
\includegraphics[trim = 0mm 0mm 1mm 0mm, clip, width=0.75\textwidth,angle=0]{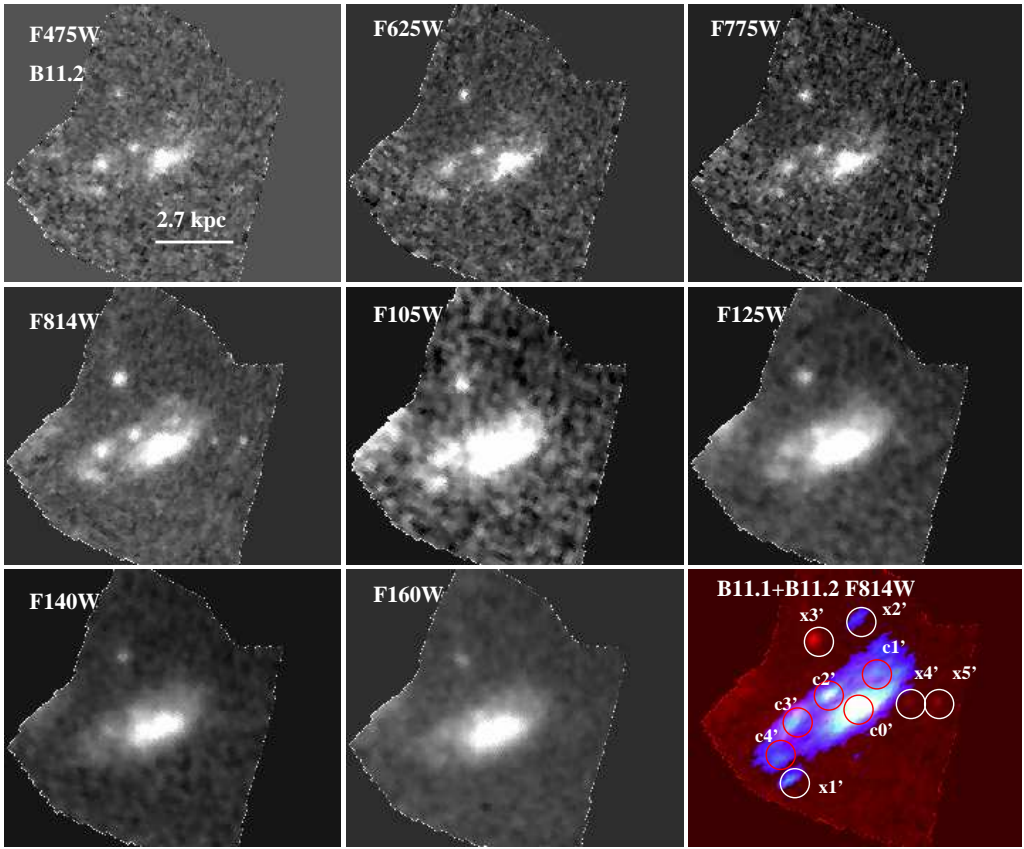}
}
\caption{ Multi-band HST images of B11.2 on the source-plane.  The last panel combines the F814W-band morphology of B11.1 and B11.2. The clumps that
are associated with the spiral host are marked in red circles whereas the interlopers are marked in white circles  (see also Section~\ref{subsec:hstmor} and Figure~\ref{fig:fig7}).
\label{fig:figa2}}
\end{figure*}

\appendix
\label{s:appendix}
We show the source-plane reconstructed multi-wavelength morphology from image B11.1 and B11.2 in Figure~\ref{fig:figa1} and Figure~\ref{fig:figa2}.
\vspace{0.3cm}

\end{document}